\PassOptionsToPackage{unicode}{hyperref}
\PassOptionsToPackage{hyphens}{url}
\PassOptionsToPackage{dvipsnames,svgnames,x11names}{xcolor}
\documentclass[
  12pt]{article}

\usepackage{amsmath,amssymb}
\usepackage{iftex}
\ifPDFTeX
  \usepackage[T1]{fontenc}
  \usepackage[utf8]{inputenc}
  \usepackage{textcomp} 
\else 
  \usepackage{unicode-math}
  \defaultfontfeatures{Scale=MatchLowercase}
  \defaultfontfeatures[\rmfamily]{Ligatures=TeX,Scale=1}
\fi
\usepackage{lmodern}
\ifPDFTeX\else  
\fi
\IfFileExists{upquote.sty}{\usepackage{upquote}}{}
\IfFileExists{microtype.sty}{
  \usepackage[]{microtype}
  \UseMicrotypeSet[protrusion]{basicmath} 
}{}
\makeatletter
\@ifundefined{KOMAClassName}{
  \IfFileExists{parskip.sty}{%
    \usepackage{parskip}
  }{
    \setlength{\parindent}{0pt}
    \setlength{\parskip}{6pt plus 2pt minus 1pt}}
}{
  \KOMAoptions{parskip=half}}
\makeatother
\usepackage{xcolor}
\makeatletter
\ifx\paragraph\undefined\else
  \let\oldparagraph\paragraph
  \renewcommand{\paragraph}{
    \@ifstar
      \xxxParagraphStar
      \xxxParagraphNoStar
  }
  \newcommand{\xxxParagraphStar}[1]{\oldparagraph*{#1}\mbox{}}
  \newcommand{\xxxParagraphNoStar}[1]{\oldparagraph{#1}\mbox{}}
\fi
\ifx\subparagraph\undefined\else
  \let\oldsubparagraph\subparagraph
  \renewcommand{\subparagraph}{
    \@ifstar
      \xxxSubParagraphStar
      \xxxSubParagraphNoStar
  }
  \newcommand{\xxxSubParagraphStar}[1]{\oldsubparagraph*{#1}\mbox{}}
  \newcommand{\xxxSubParagraphNoStar}[1]{\oldsubparagraph{#1}\mbox{}}
\fi
\makeatother

\usepackage{longtable,booktabs,array}
\usepackage{calc} 
\usepackage{etoolbox}
\makeatletter
\patchcmd\longtable{\par}{\if@noskipsec\mbox{}\fi\par}{}{}
\makeatother
\IfFileExists{footnotehyper.sty}{\usepackage{footnotehyper}}{\usepackage{footnote}}
\makesavenoteenv{longtable}
\usepackage{graphicx}
\makeatletter
\def\maxwidth{\ifdim\Gin@nat@width>\linewidth\linewidth\else\Gin@nat@width\fi}
\def\maxheight{\ifdim\Gin@nat@height>\textheight\textheight\else\Gin@nat@height\fi}
\makeatother
\setkeys{Gin}{width=\maxwidth,height=\maxheight,keepaspectratio}
\makeatletter
\def\fps@figure{htbp}
\makeatother

\makeatletter
\@ifpackageloaded{caption}{}{\usepackage{caption}}
\AtBeginDocument{%
\ifdefined\contentsname
  \renewcommand*\contentsname{Table of contents}
\else
  \newcommand\contentsname{Table of contents}
\fi
\ifdefined\listfigurename
  \renewcommand*\listfigurename{List of Figures}
\else
  \newcommand\listfigurename{List of Figures}
\fi
\ifdefined\listtablename
  \renewcommand*\listtablename{List of Tables}
\else
  \newcommand\listtablename{List of Tables}
\fi
\ifdefined\figurename
  \renewcommand*\figurename{Figure}
\else
  \newcommand\figurename{Figure}
\fi
\ifdefined\tablename
  \renewcommand*\tablename{Table}
\else
  \newcommand\tablename{Table}
\fi
}
\@ifpackageloaded{float}{}{\usepackage{float}}
\floatstyle{ruled}
\@ifundefined{c@chapter}{\newfloat{codelisting}{h}{lop}}{\newfloat{codelisting}{h}{lop}[chapter]}
\floatname{codelisting}{Listing}

\makeatother
\makeatletter
\@ifpackageloaded{caption}{}{\usepackage{caption}}
\@ifpackageloaded{subcaption}{}{\usepackage{subcaption}}
\makeatother

\ifLuaTeX
  \usepackage{selnolig}  
\fi
\usepackage[]{natbib}
\usepackage{bookmark}
\IfFileExists{xurl.sty}{\usepackage{xurl}}{} 
\hypersetup{
  pdftitle={Title},
  pdfauthor={Author 1; Author 2},
  pdfkeywords={3 to 6 keywords, that do not appear in the title},
  colorlinks=true,
  linkcolor={blue},
  filecolor={Maroon},
  citecolor={Blue},
  urlcolor={Blue},
  pdfcreator={LaTeX via pandoc}}

\newcommand{\anon}{1}

\newcommand\fnote[1]{\captionsetup{font=footnotesize}\caption*{#1}}

\begin{document}

\def\spacingset#1{\renewcommand{\baselinestretch}%
{#1}\small\normalsize} \spacingset{1}


\if1\anon
{
  \title{\bf A Multinomial Probit Model\\ for\\ Asymmetric Choice Responses}
  \author{Cash Looi\thanks{
    Correspondence to: Department of Econometrics \& Business Statistics, Monash University, Clayton VIC 3800, Australia, e-mail: \textsf{cash.looi@monash.edu}}\hspace{.2cm}\\
    Rub\'en Loaiza-Maya \\
    Didier Nibbering\\
    Department of Econometrics and Business Statistics, Monash University}
  \maketitle
} \fi

\if0\anon
{
  \bigskip
  \bigskip
  \bigskip
  \begin{center}
    {\LARGE\bf Title}
\end{center}
  \medskip
} \fi

\bigskip
\begin{abstract}
Standard multinomial probit (MNP) models specify symmetric latent utility distributions, implying that choice probabilities respond symmetrically to positive and negative covariate shifts of the same magnitude. This restriction is often implausible in empirical choice settings and can lead to misleading elasticity and substitution predictions. 
We propose a skewed multinomial probit (SMNP) model that captures asymmetric choice responses by specifying a multivariate skew-normal distribution for the latent utilities. The model preserves the flexible substitution patterns of the MNP framework, introduces alternative-specific skewness parameters, and nests the standard MNP model when skewness is zero. 
Introducing skewness creates identification and computational challenges because the skewness parameters interact with the MNP scale normalization and disrupt the conditional Gaussian updating structure used in Bayesian MNP estimation. We address these challenges through a covariance reparameterization that enforces identification and positive definiteness by construction, interpretable priors on the identified parameter space, and a double data-augmentation scheme that yields a Metropolis-Hastings within Gibbs sampler. Numerical experiments and applications to consumer choice data show that SMNP recovers asymmetric choice responses, improves probabilistic prediction, and produces economically meaningful differences in price elasticities and substitution patterns.
\end{abstract}

\noindent%
{\it Keywords:} 
Skew-normal distribution; discrete choice; data augmentation; price elasticities; substitution patterns
\vfill

\newpage
\spacingset{1.8} 

\section{Introduction}
Discrete choice models often impose symmetric choice responses: they imply that a positive change in an attribute of a choice alternative affects choice probabilities by the same magnitude as an equivalent negative change. This restriction arises when a linear latent utility function is combined with a symmetric error distribution. However, in many decision-making settings, this symmetry is difficult to justify. People may react differently to increases and decreases in the same attribute, so the probability of choosing an alternative may change more strongly in one direction than in the other.

The idea of asymmetric choice responses has a long foundation in behavioral economics. Prospect Theory argues that agents respond more strongly to perceived losses than to perceived gains \citep{kahnemann1979prospect}. This has motivated a large empirical literature on asymmetric responses to price changes, especially in economics and marketing. Recent studies document loss aversion in settings ranging from child health care \citep{iizuka2023asymmetric} and soft drinks \citep{biondi2020between} to ketchup and milk \citep{caputo2018choice}. Capturing such asymmetric choice responses matters for governments that evaluate welfare policies and for firms that design pricing strategies.

This paper proposes a skewed multinomial probit (SMNP) model that allows for asymmetric choice responses. The model builds on the multinomial probit (MNP) framework, which is well suited to choice applications because it allows the latent utilities of different alternatives to be correlated. These correlations capture general substitution patterns among alternatives and avoid the independence of irrelevant alternatives restriction imposed by other choice models \citep{hausman1978conditional}. However, the standard MNP model still imposes symmetric choice responses by specifying a multivariate normal distribution for the latent utilities conditional on observed covariates. We relax this restriction by replacing the multivariate normal distribution with a multivariate skew-normal distribution. The resulting SMNP model introduces alternative-specific skewness parameters, allowing both the degree and direction of asymmetry to vary across alternatives in the choice set.

Introducing skewness into the MNP framework creates new identification challenges, which we address with a novel reparameterization of the implied covariance matrix. MNP models already require a scale normalization in the covariance matrix because the observed choices are invariant to rescalings of the latent utilities. In the SMNP model, this normalization interacts with the skewness parameters because the covariance structure depends on both the scale and the skewness parameters in the multivariate skew-normal distribution. As a result, one cannot freely estimate the skewness parameters and covariance matrix. We solve this problem by developing a reparameterization that fixes the scale and ensures a positive definite covariance matrix.

We address the computational challenges in the estimation of the SMNP model by developing Bayesian inference based on data augmentation and Markov chain Monte Carlo (MCMC) sampling. In the MNP model, likelihood evaluation is costly because choice probabilities involve multivariate integrals. Bayesian estimation reduces this burden by augmenting the model with latent utilities and sampling them inside a MCMC scheme \citep{albert1993bayesian}. This strategy relies on the conditional normality of the latent utilities and therefore does not directly transfer to SMNP. We address this problem by introducing a second layer of augmentation based on the conditionally Gaussian representation of the multivariate skew-normal distribution \citep{fruhwirth2010bayesian}. The double data augmentation yields closed-form updates for all parameters but one, which we update with a single Metropolis-Hastings step.

This paper contributes to a large literature that captures asymmetric choice responses by modifying the observable component of utility in discrete choice models. Typically, researchers split a focal attribute, such as price, into gains and losses relative to a reference point \citep{kalwani1990price, krishnamurthi1992asymmetric, hardie1993modeling, briesch1997comparative,mazumdar2005reference, caputo2018choice, biondi2020between}. The model then detects asymmetry when the coefficient on losses differs from the coefficient on gains. \citet{bhat2012new} introduce asymmetry through random coefficients with a skew-normal distribution. Alternatively, asymmetric responses can be generated by using more flexible predictor functions, such as nonlinear transformations, polynomial terms, or splines. Instead of imposing asymmetry through observed attributes, we allow asymmetry to arise from the conditional distribution of the latent utilities. This parsimonious way of capturing asymmetric choice responses does not require reference points, random coefficients, or predictor transformations.

A second strand of literature introduces asymmetry through the error distribution or link function, primarily for binary response data: \citet{chen1999new, bazan2010framework,kim2001bayesian,kim2002binary,kim2008flexible,zhang2023tractable}. However, choice applications often involve more than two alternatives, and extending to multinomial outcomes creates additional identification and computational challenges. Recent work has derived Bayesian conjugacy properties for a class of skewed MNP models assuming a fixed scale matrix \citep{fasano2022class, anceschi2023bayesian,karling2024conjugacy}. Conditioning on a fixed scale matrix is restrictive in applications, where the dependence across latent utilities is unobserved and often central for capturing substitution patterns. Moreover, these results do not deliver joint inference over all model parameters, nor do they address the identification problems that arise when skewness is introduced. We fill this gap by developing a fully identified SMNP model and an empirically implementable Bayesian inference method. 

We illustrate the practical relevance of our method with two numerical experiments and two empirical applications. The numerical experiments compare the SMNP and MNP models under data generating processes with asymmetric and symmetric choice responses. We find that SMNP recovers asymmetric choice responses and improves predictive accuracy relative to the MNP. When choice responses are symmetric, SMNP performs comparably to the MNP model. The empirical applications to laundry detergent and ketchup purchases show that asymmetries matter in practice. The SMNP detects asymmetry in both applications, these asymmetries change the implied own- and cross-price elasticities, and improves prediction. The MNP understates price sensitivity for detergent brands, and misallocates substitution across ketchup brands. These results indicate that the SMNP model can help practitioners make better pricing and substitution decisions.

The remainder of this article is structured as follows. Section \ref{skewed} introduces the SMNP model specification, while Section \ref{bayes} covers aspects of Bayesian estimation pertaining to data augmentation, prior specification, and the proposed MCMC sampler. Section \ref{sim} evaluates the proposed model through the numerical simulation studies. Section \ref{empirical} then applies the proposed method to real consumer datasets, followed by a discussion in Section \ref{discuss}. 

\section{Skewed multinomial probit model}\label{skewed}  
\subsection{Model specification}\label{model}  
Let $Y_i\in \{1,2,\dots,J+1\}$ be an observed discrete choice with $J+1$ the number of choice alternatives, and $i=1,2,\dots,N$ with $N$ the number of individuals. It is assumed that each $Y_i$ is associated with a $J\times1$ vector of latent utilities $Z_{i}=(z_{i1}, z_{i2},\dots, z_{iJ})^\top$: 
\begin{equation}\label{eq:Y}
Y_i=
\begin{cases}
J+1,\quad  &\text{if }\ \max(Z_i)<0 ,\\
j,\quad  &\text{if }\ z_{ij}=\max(Z_i)>0,\\
\end{cases}
\end{equation}
where $\max(Z_i)$ denotes the largest value among the elements in the vector $Z_i$. Category $J+1$ is referred to as the base category. The latent utilities are modeled as
\begin{equation}\label{eq:Zmsn}
   Z_i=X_i\beta+\varepsilon_i, \quad \quad \varepsilon_i \sim SN_J(0,{\Sigma},{\alpha}),
\end{equation}
where $X_i$ represents the $J \times k$ regressor matrix and $\beta$ is a $k \times 1$ vector of coefficients. The $J\times 1$ disturbance vector $\varepsilon_i$ follows a $J-$variate skew normal distribution defined in \cite{azzalini1996multivariate} and \cite{azzalini1999statistical}, with $\Sigma$ a $J\times J$ positive definite scale matrix and ${\alpha} = (\alpha_1, \dots, \alpha_J)^\top \in \mathbb{R}^J$ a vector of shape parameters governing asymmetry. 

The model specified in \eqref{eq:Y} and \eqref{eq:Zmsn} considers only $J$ utilities for each choice $Y_i$, while there are $J+1$ choice alternatives. This reduction to a $J$-dimensional space is to address the location identification problem that follows if a unique utility is specified for each choice alternative \citep{bunch1991estimability}. In addition, we have to address the scale identification problem: $Y_i(cZ_i) = Y_i(Z_i)$ for any $c \in \mathbb{R}^+$. We follow the common approach by fixing the first diagonal element of $\Sigma$ to one.

The regressor matrix typically includes $J$ alternative-specific intercepts, a $k_d$-dimensional vector $x_{i,d}$ of individual-specific covariates, and a $(J+1)\times k_a$ matrix $x_{i,a}$ of $k_a$ alternative-specific covariates. Therefore, $k=J + Jk_d+k_a$ and $X_i$ can be written as
\begin{equation}\label{eq:X}
   X_i = \left[I_J \quad x_{i,d}^\top \otimes I_J \quad Tx_{i,a}    \right],
\end{equation}
where $I_J$ denotes the $J\times J$ identity matrix, $ T = \bigl[\, I_J \;\; -\mathbf{1}_J \,\bigr]$,  and $\mathbf{1}_J$ is a $J \times 1$ vector of ones. Specifically, the transformation matrix $T$ constructs the differences in the alternative-specific covariates between each alternative and the $(J+1)$-th alternative.

\subsection{Asymmetric choice responses} %
Since the multivariate skew normal distribution is closed under affine transformations \citep{Azzalini_2013}, the latent utilities satisfy $
Z_i \sim SN_J(X_i\beta,{\Sigma},{\alpha})$. When the shape parameter vector is a $J-$dimensional vector of zeros, such that $\alpha=0_J$, the distribution reduces to a $J$-variate normal distribution with mean $X_i\beta$ and covariance matrix $\Sigma$. Thus, the SMNP model encompasses the multinomial probit model (MNP) as a special case. 

The shape vector $\alpha$ governs asymmetry in the multivariate distribution of the latent utilities, but its components may be difficult to interpret. We therefore use the skewness-parameter vector $\delta = (1+\alpha^\top\Sigma\alpha)^{-1/2}\Sigma\alpha$, whose elements directly determine the direction and magnitude of skewness in the marginal latent utilities. Throughout the paper, we refer to $\delta$ as the skewness parameter vector. Since $\Sigma$ is positive definite, $\alpha=0_J$ if and only if $\delta=0_J$.

Skewness in the conditional distribution of the latent utilities changes how choice probabilities respond to covariate shifts. Under the symmetric MNP model, positive and negative changes in a covariate of the same magnitude generate symmetric probability responses around a given reference point. The SMNP model relaxes this restriction. It allows the probability of choosing an alternative to react more strongly to a change in one direction than to an equally sized change in the opposite direction.

\begin{figure}[tb!]
\caption{Asymmetric latent utilities and choice probability responses}
\centering
\includegraphics[width=\textwidth]{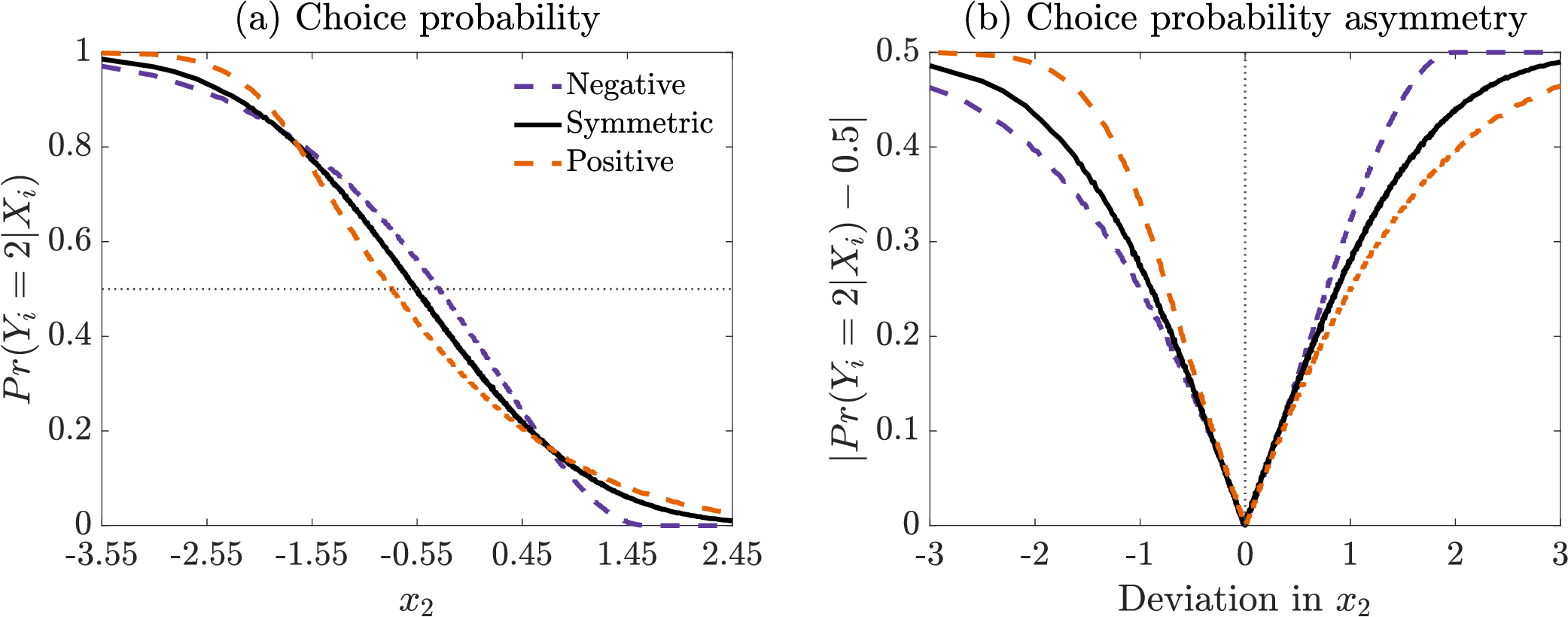}
\fnote{This figure shows the impact of asymmetric latent utilities on choice probabilities. Panel (a) shows the choice probability for category $2$ as a function of its alternative-specific covariate $x_2$ for negatively skewed ($\delta=(0,-1.437)^\top$, dashed purple), symmetric ($\delta=(0,0)^\top$, solid black), and positively skewed ($\delta=(0,1.437)^\top$, dashed orange) latent utilities. Panel (b) shows the absolute deviation of the choice probabilities from the reference probability $0.5$ as a function of the deviations in $x_2$. Probabilities are simulated using \eqref{eq:Y} and \eqref{eq:Zmsn} with $\beta = (0,0,0.7)^\top$, $\Sigma=[1\,0.5;0.5\,1]$ in the MNP, and $x_2$ a standard normal distribution.}
\label{fig:cpplot}
\end{figure}

\noindent\textbf{Example:} Figure~\ref{fig:cpplot} illustrates this mechanism in a three-alternative choice setting. We compare three specifications that have the same conditional mean and covariance for the latent utilities but differ in skewness: a negatively skewed SMNP (dashed purple line), a symmetric MNP (solid black line), and a positively skewed SMNP (dashed orange) specification. Panel (a) shows that the skewness parameter changes the shape of the choice probability curve. Panel (b) makes the asymmetry explicit by plotting the absolute deviation of the choice probability from 0.5 around the corresponding reference value of the covariate. 

Panel (b) shows that, under the symmetric MNP model, equal positive and negative deviations in the covariate produce equal absolute changes in choice probability. This may be unrealistic in many empirical choice settings. For instance, positive and negative price changes may have different effects on purchase probabilities. In the SMNP model, these deviations produce different changes, generating asymmetric choice responses. For instance, with negatively (positively) skewed latent utilities, negative deviations in the covariate result in smaller (larger) changes in choice probability relative to changes as a result of positive deviations in the covariate. These asymmetries may reflect loss aversion, brand loyalty, or switching costs in consumer choice settings.

\section{Bayesian inference for the identified SMNP model} \label{bayes}
Estimating the SMNP model requires addressing two challenges. The first is computational. To avoid evaluation of the multivariate integrals in the likelihood function, Bayesian estimation augments the MNP model with latent utilities. In the SMNP model, however, augmenting only with the latent utilities does not deliver the same conjugate Gaussian updating structure. The second challenge is identification. As discussed in Section~\ref{model}, MNP models require a scale normalization. In the SMNP model, this normalization interacts with the skewness parameters. Hence, the skewness vector and the scale matrix cannot be sampled independently. 

We address both challenges jointly. We first introduce a second layer of augmentation that gives the skew-normal distribution a conditionally Gaussian representation. We then impose the MNP scale normalization in this augmented parameterization. This yields an identified model and a tractable Metropolis-Hastings within Gibbs sampler.

\subsection{Double data augmentation}
Let $Z=(Z_1^\top,\dots,Z_N^\top)^\top$ and $X=(X_1^\top,\dots,X_N^\top)^\top$, and let $\theta$ denote the parameter vector. If we augment the observed choices only with the latent utilities, the posterior is
\begin{equation}\label{eq:post1}
p(\theta, Z\mid Y,X) \propto  \prod^N_{i=1} p(Y_i\mid Z_i)p(Z_i\mid X_i, \theta)p(\theta),
\end{equation}
where $ p(Y_i\mid Z_i)$ is implied by \eqref{eq:Y} and $p(\theta)$ denotes the prior distribution \citep{albert1993bayesian}. For the standard MNP model, this representation leads to conditionally normal latent utilities, and Gibbs sampling is directly applicable to the posterior in \eqref{eq:post1}. For the SMNP model, $p(Z_i\mid X_i,\theta)$ is skew-normal, so this augmentation alone does not admit conjugate conditional posteriors for the latent utilities $Z_i$. 

We obtain a conditional normal representation in the skew-normal utilities by using the stochastic representation of the multivariate skew-normal distribution, see \citet{fruhwirth2010bayesian}. In particular, we write
 \begin{equation}\label{eq:equil}
\begin{split} 
p(Z_i\mid X_i, \theta) 
& = 2\phi_J(Z_i;X_i\beta; {\Sigma})\Phi({\alpha}^\top({Z_i}-X_i\beta))\\
&= \int_{0}^\infty \phi_J(Z_i;X_i\beta+\delta w_i,\Sigma_u)p(w_i) dw_i,
\end{split}
\end{equation}
where $\phi_{J}\left(Z;\mu,\Sigma\right)$ denotes the $J$-variate normal density with mean $\mu$ and covariance matrix $\Sigma$, $\Phi(\cdot)$ the univariate standard normal cumulative distribution function, $p(w_i) = 2\phi_1(w_i;0,1)\mathbb{I}(w_i>0)$ the univariate truncated standard normal density, and $\Sigma_u = \Sigma-\delta\delta^\top$. From here onward, we parameterize the model in terms of $\delta$ and $\Sigma_u$,  noting that $\alpha= (1-\delta^\top\Sigma^{-1}\delta)^{-1/2}\Sigma^{-1}\delta$ and $\Sigma = \Sigma_u+\delta\delta^\top$.   

The representation in \eqref{eq:equil} allows for a second layer of data augmentation using $w_i$. The resulting doubly augmented posterior becomes 
\begin{equation}\label{eq:augpost} 
p(\theta, Z, w \mid Y,X) \propto \prod^N_{i=1}p(Y_i\mid Z_i)\phi_J(Z_i;X_i\beta+\delta w_i,\Sigma_u)p(w_i)p(\theta),
\end{equation}
where $w=(w_1,\dots,w_N)^\top$. Conditional on $w$, the latent utilities are Gaussian and a conjugate conditional posterior for $Z_i$ is available. Hence, this second augmentation step turns the skew-normal latent utility model into a conditionally Gaussian regression model, which yields closed-form conditional updates for most model components.

\subsection{Scale identification in the doubly augmented parameterization} \label{identification}
The double augmentation introduces the covariance matrix $\Sigma_u$, but the scale normalization $\Sigma_{11}=1$ discussed in Section~\ref{model} applies to the skew-normal scale matrix $\Sigma$. Since $\Sigma_u= \Sigma-\delta\delta^\top$, the scale restriction imposes that the first element of $\Sigma_u$ is fixed at $1-\delta^2_1$. This restriction shows why the covariance matrix and skewness vector cannot be treated as independent parameters. Sampling an unrestricted positive definite matrix for $\Sigma_u$ would generally violate the scale normalization of the SMNP model.

We impose the restriction directly by parameterizing $\Sigma_u$ as 
\begin{equation}\label{eq:sig_u}
\Sigma_u=
\begin{bmatrix}
1-\delta_1^2 & \gamma^\top\\
\gamma &  \psi+ \dfrac{\gamma\gamma^\top}{1-\delta_1^2} 
\end{bmatrix},
\end{equation} 
where $\gamma$ is a $(J-1)$-dimensional vector and $\psi$ is a $(J-1)$-dimensional positive definite square matrix. If there is no skewness and $\delta_1=0$, this representation coincides with the parametrization of the covariance matrix in the MNP model of \cite{mcculloch2000bayesian}. 

The parameterization in \eqref{eq:sig_u} is key for identification in the SMNP model which allows for $\delta_1\neq 0$. If $|\delta_1|<1$ and $\psi$ is positive definite, the Schur complement of $(1-\delta_1^2)$ in $\Sigma_u$ is exactly $\psi$, so $\Sigma_u$ is positive definite. Moreover, the implied scale matrix satisfies $\Sigma_{11}=(1-\delta_1^2)+\delta_1^2=1$. Thus, every posterior draw generated under this parameterization automatically satisfies the scale normalization and remains inside the positive definite parameter space.
The identified parameter vector is now $\theta = (\beta^\top,\delta^\top,\gamma^\top,\text{vec}(\psi)^\top)^\top$. Given draws of $\delta_1$, $\gamma$, and $\psi$, we construct $\Sigma_u$ using \eqref{eq:sig_u} and recover $\Sigma=\Sigma_u+\delta\delta^\top$. This implies that $\delta_1$ enters all expressions of the augmented likelihood that depend directly on $\Sigma_u$. Moreover, $\delta_1$ and $\psi$ require priors that ensure $|\delta_1|<1$ and the strictly positive definiteness of $\psi$.

\subsection{Prior specification under the identified parameterization}\label{prior}
We place priors directly on the identified parameter vector $\theta = (\beta^\top,\delta^\top,\gamma^\top,\text{vec}(\psi)^\top)^\top$. We use a multivariate normal prior on the model coefficients, $\beta \sim \mathcal{N}_k(0_k,10I_k)$. For the asymmetry parameter $\delta$, we use a multivariate normal prior with a truncation on $\delta_1$: $\delta \sim \mathcal{N}_J(0_J,\tau_\delta I_J)\,\mathbb{I}(|\delta_1|<c)$, where $c<1$ is a numerical bound that enforces the scale-identification constraint. We set $c=0.995$. Centering this prior at zero is important because $\delta=0_J$ corresponds to the standard MNP model. Hence, the prior does not force asymmetry into the latent utilities. The hyperparameter $\tau_\delta$ controls how much prior mass is assigned to asymmetric latent utility distributions.

For the covariance parameters, we specify $\gamma \sim \mathcal{N}_{J-1}(B_\gamma,\tau_\gamma I_{J-1})$ and $\psi \sim \mathcal{IW}_{J-1}(J+3,V)$. Because $\Sigma=\Sigma_u+\delta\delta^\top$, the prior on $\delta$ affects the implied prior distribution of the scale matrix $\Sigma$. We therefore choose the hyperparameters of the priors for $\gamma$ and $\psi$ jointly with the prior for $\delta$, so that the implied prior mean of $\Sigma$ is the equicorrelated matrix $\mathbb{E}(\Sigma) = \frac{1}{2}\,1_J1_J^\top + \frac{1}{2}\,I_J$ proposed in \cite{geweke1994alternative}. This construction centers the model on a familiar MNP covariance structure while allowing departures from symmetry through $\delta$.

The prior specification balances two goals. First, it preserves the standard MNP model as a meaningful special case by placing prior mass around $\delta=0_J$. Second, it leaves enough prior variation in $\delta$ to allow the data to reveal skewness when asymmetric choice responses are present. Values of $\tau_\delta$ that are too small shrink the model toward the symmetric MNP specification, while values that are too large can force dependence in the latent utilities to be captured through skewness rather than through the covariance structure. Appendix~\ref{prioreliccit} derives the hyperparameter restrictions that ensure the implied prior for $\Sigma_u$ remains positive definite and gives the resulting choices of $\tau_\delta$, $B_\gamma$, $\tau_\gamma$, and $V$.

\subsection{Posterior simulation} 
The double augmentation and identified parameterization yield a tractable Metropolis-Hastings within Gibbs sampler. Conditional on $Z$ and $w$, the model is Gaussian, so all parameters except $\delta_1$ have standard full conditional distributions. This parameter enters both the skewness term  and the normalized covariance matrix $\Sigma_u$, because $(\Sigma_u)_{11}=1-\delta_1^2$. As a result, its full conditional distribution is not available in closed form.

We update $\delta_1$ using a univariate Metropolis-Hastings step. To respect the constraint $|\delta_1|<c$, we transform $\delta_1$ to the real line and construct a proposal distribution from a Laplace approximation to the transformed conditional posterior. All other parameters are updated using Gibbs steps. The sampler therefore isolates the nonconjugacy created by scale identification under skewness in a single one-dimensional Metropolis-Hastings update.

Starting from initial values $\{\delta_{1:J}^{(0)},w^{(0)},\gamma^{(0)},\psi^{(0)},Z^{(0)},\beta^{(0)}\}$, we iterate the following steps for $r=1,2,\dots,R$:
\begin{description}
    \item[\textbf{Step 1:}] Generate $\delta_1^{(r)} \sim p(\delta_1 \mid\delta_{1:J}^{(r-1)}, w^{(r-1)},\gamma^{(r-1)},\psi^{(r-1)}, Z^{(r-1)}, \beta^{(r-1)},Y,X)$.
    \item[\textbf{Step 2:}] Generate $\delta_{2:J}^{(r)} \sim p(\delta_{2:J} \mid\delta_{1}^{(r)}, w^{(r-1)},\gamma^{(r-1)},\psi^{(r-1)}, Z^{(r-1)}, \beta^{(r-1)},Y,X)$.
    \item[\textbf{Step 3:}] Generate $w^{(r)} \sim p(w \mid \delta^{(r)},\gamma^{(r-1)},\psi^{(r-1)}, Z^{(r-1)},\beta^{(r-1)},Y,X)$.
    \item[\textbf{Step 4:}] Generate $\gamma^{(r)} \sim p(\gamma \mid \delta^{(r)}, w^{(r)},\psi^{(r-1)}, Z^{(r-1)},\beta^{(r-1)},Y,X)$.
    \item[\textbf{Step 5:}] Generate $\psi^{(r)} \sim p(\psi \mid \delta^{(r)}, w^{(r)},\gamma^{(r)}, Z^{(r-1)},\beta^{(r-1)},Y,X)$.
    \item[\textbf{Step 6:}] Generate $Z^{(r)} \sim p(Z \mid \delta^{(r)}, w^{(r)},\gamma^{(r)},\psi^{(r)},\beta^{(r-1)},Y,X)$.
    \item[\textbf{Step 7:}] Generate $\beta^{(r)} \sim p(\beta \mid \delta^{(r)}, w^{(r)},\gamma^{(r)},\psi^{(r)},Z^{(r)},Y,X)$.
\end{description}

At each iteration, $\Sigma_u^{(r)}$ and $\Sigma^{(r)}$ are deterministic functions of $\delta^{(r)}$, $\gamma^{(r)}$, and $\psi^{(r)}$. The sampler therefore produces posterior draws that satisfy both positive definiteness and the scale normalization. Appendix~\ref{MCMC} provides the full conditional distributions and further implementation details.

\subsection{Predictive distribution and scoring rules}
In discrete choice models, the model parameters are generally not directly interpretable, and interest usually focuses on the predictive distribution of $Y_i$:
\begin{equation}
Pr(Y_i =j\mid X_i,Y,X)= \int \int p(Y_i =j\mid Z_i)p(Z_i\mid X_i, \theta)p(\theta \mid Y,X) \ dZ_i \ d\theta.
\end{equation}
Note that the parameter vector $\theta$ for SMNP and MNP represent different sets of model parameters. As this integral is not available in closed form, we evaluate it using MCMC draws. For each retained posterior draw $\theta^{(m)}$ with $m=1,\dots, M$, we generate $Z_i^{(m)} \sim p(Z_i\mid X_i, \theta^{(m)})$, and subsequently  $Y_i^{(m)} \sim  p(Y_i\mid Z_i^{(m)})$. The predictive probability is evaluated as
\begin{equation}
\hat{Pr}(Y_i =j\mid X_i,Y,X)=  \frac{1}{M} \sum^M_{m=1} \mathbb{I}(Y_i^{(m)}=j),
\end{equation} 
and the point prediction $\hat Y_i$ equals the mode of the predictive distribution.

We evaluate predictive performance using the hit rate, and the proper scoring rules such as the log score and the censored likelihood score. The hit-rate evaluates point predictions:
\begin{equation}
HR=  \frac{1}{N} \sum^N_{i=1} \mathbb{I} \left( \hat Y_i =Y_i \right).
\end{equation}
The log score evaluates the full predictive distribution:
\begin{equation}
LS=  \frac{1}{N} \sum^N_{i=1} \sum^{J+1}_{j=1} \ln(\hat{Pr}(Y_i = j \mid X_i,Y,X)) \mathbb{I}(Y_i = j).
\end{equation}
Finally, the censored likelihood score of \citet{diks2011likelihood} measures predictive accuracy for a specific category $j$: $CLS_j = $
\begin{equation}
\frac{1}{N} \sum_{i=1}^N \left[ \ln(\hat{Pr}(Y_i = j\mid X_i,Y,X)) \mathbb{I}(Y_i = j) + \ln(\hat{Pr}(Y_i \neq j \mid X_i,Y,X)) \mathbb{I}(Y_i \neq j) \right].
\end{equation}
For all three accuracy metrics, larger values indicate better predictive performance. 

\section{Numerical experiments} \label{sim}
We conduct two numerical experiments to evaluate the inferential and predictive accuracy of the SMNP relative to the MNP. The first experiment generates data from a skewed latent utility distribution, and the second experiment from a normal latent utility distribution. This allows us to assess whether the SMNP can recover asymmetric choice responses and whether the SMNP collapses to the MNP  when the skewness parameters are not needed.

\subsection{Design}\label{sec:design}

We consider a three alternative choice model with $N=20,000$ observations. We generate two datasets. The first dataset is generated from an SMNP model with skewed latent utilities, using $(\delta_0=(0.8,1.3)^\top)$. The second dataset is generated from the corresponding MNP model by setting $(\delta_0=(0,0)^\top)$. All other components of the data-generating process are kept fixed across the two experiments.

The model includes two alternative-specific intercepts and one alternative-specific regressor. We generate $x_{i,a}=(x_{i,1},x_{i,2},x_{i,3})^\top$, where $x_{i,j}\sim \text{Log Normal}(0,1)$ independently across observations and alternatives. We then use $\log x_{i,a}$ as the covariate, so that the regressor can be interpreted as log price. The true coefficient vector is $\beta_0=(-0.5,-0.8,-0.3)^\top$, where the first two elements are alternative-specific intercepts, and the third element is the price coefficient. The diagonal elements of the true scale matrix $\Sigma_0$ are $1$ and $1.79$, and the off-diagonal equals $1.04$. Given these parameters, we generate data from \eqref{eq:Y} and \eqref{eq:Zmsn}. 

For each dataset, we randomly assign half of the observations to the training sample and half to the test sample. We estimate both the SMNP and MNP models on the training sample. For the SMNP model, we use the prior specification in Section~\ref{prior}. For the MNP model, we use the same prior for $\beta$ and the covariance prior of \citet{mcculloch2000bayesian}, centered on an equicorrelated matrix with off-diagonal elements equal to 0.5. We run the SMNP and the MNP samplers for 1,000,000 iterations, discarding the first half of each chain as burn-in. For the construction of predictive probabilities and scoring rules, we thin the retained draws to 10,000 posterior draws.

\subsection{Results with asymmetric choice responses}
\begin{figure}[tb!]
\centering
\includegraphics[width=\textwidth]{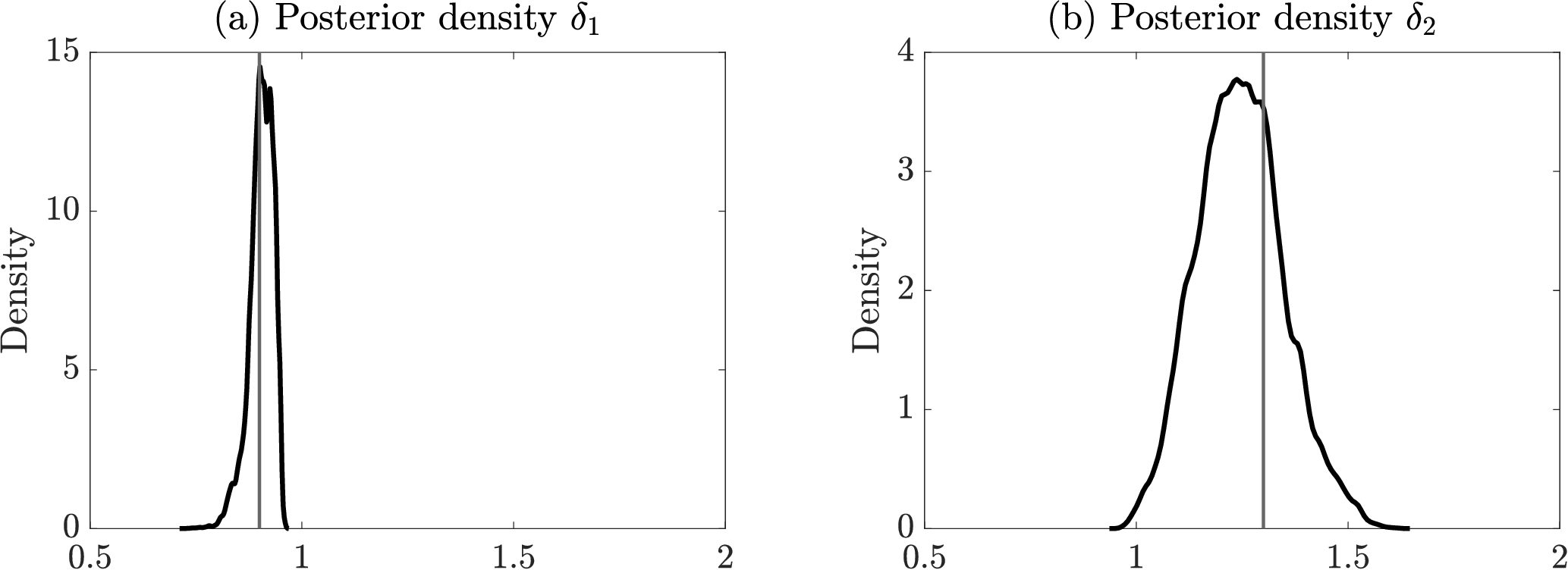}
\caption{Posterior densities $\delta_1$ and $\delta_2$ in the SMNP model for the experiment with skewed latent utilities. The vertical lines represent true parameter values.}
\label{fig:idendeltaMSN}
\end{figure}

We first examine the experiment in which the data are generated from the SMNP model. Figure~\ref{fig:idendeltaMSN} shows the posterior densities of the two skewness parameters in the SMNP model. The true parameter values, indicated by the vertical lines, are covered with high posterior probability mass, while the posteriors do not allocate any density mass at zero. This shows that the SMNP model captures the asymmetry in the latent utility distribution. 

\begin{figure}[tb!]
\centering
\includegraphics[width=\textwidth]{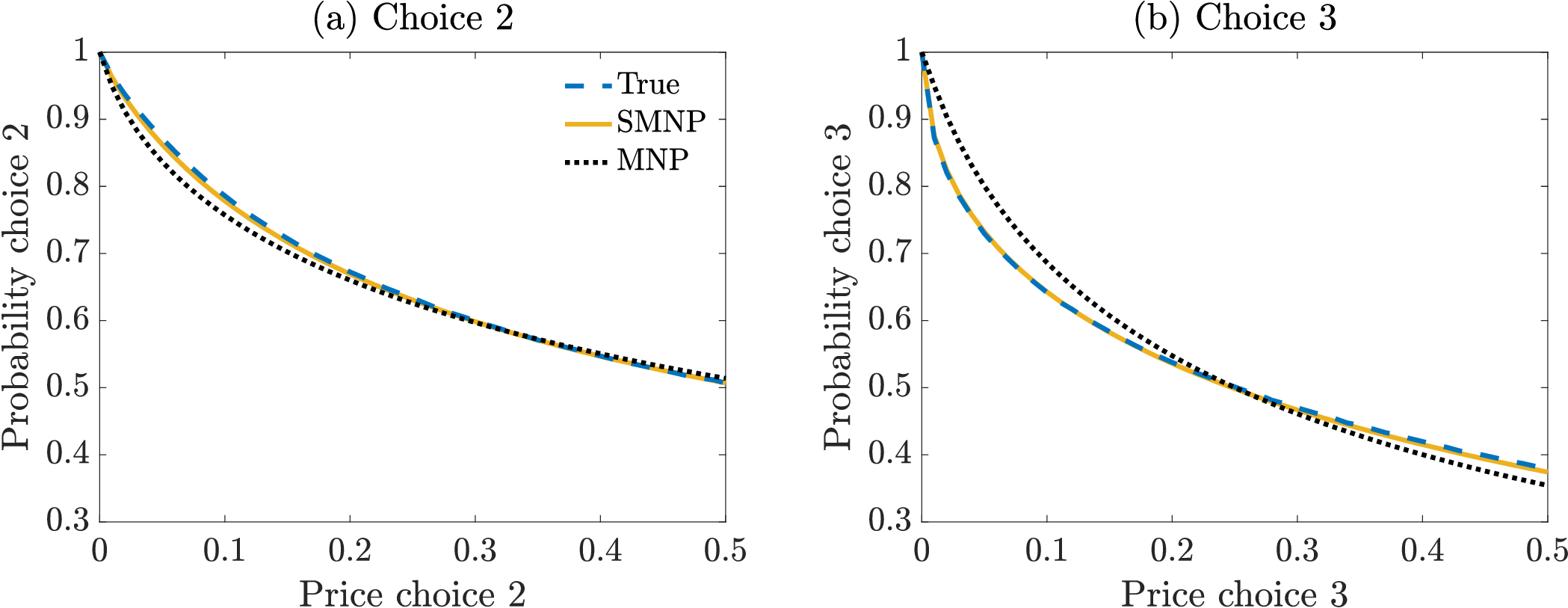}
\caption{Posterior choice probabilities for the experiment with skewed latent utilities. 
The probability of choice $2$ and choice $3$ are a function of their price, with the prices of the other choices fixed at their mean. The solid yellow and dotted black lines represent the estimated choice probabilities using the SMNP and MNP models, respectively. The dashed blue line indicates the true choice probability.}
\label{fig:choiceprob}
\end{figure}

Next, we compare the posterior choice probabilities from the SMNP and MNP models with the true choice probabilities implied by the data-generating process. Figure~\ref{fig:choiceprob} shows choice probabilities for an increasing price level for a specific alternative, with the prices of the other alternatives held constant at their mean. The choice probabilities from SMNP, MNP, and data generating process (DGP) are represented by the solid yellow, dotted black, and dashed blue lines, respectively. The SMNP model closely tracks the true probability curves. The MNP model, in contrast, displays systematic deviations, especially in the tails of the price range. For instance, Panel (a) demonstrates that the MNP model underestimates the true choice probabilities for prices below $0.2$. In Panel (b), the MNP model exhibits bias at both ends of the price level. These differences show that the symmetric latent utility specification distorts the implied substitution patterns when the true latent utilities are skewed.

We examine how these discrepancies in choice probabilities translate into differences in predictive accuracy across the models. Panel A of Table~\ref{tab:MSNpredictive} reports the predictive performance under the asymmetric DGP, comparing the MNP and SMNP models against predictions from the true DGP, hereafter referred to as the oracle. In-sample, the SMNP model attains a higher hit rate, and performs better on the log score and all censored likelihood scores. The log-score and $CLS_3$ of SMNP are close to these metrics for the oracle, whereas the MNP model produces statistically significantly lower values according to the Giacomini--White test. Out-of-sample, the SMNP model also improves both point and probabilistic prediction. The SMNP log score and censored likelihood scores are also close to the oracle values, whereas the MNP model differs significantly from the oracle for the log score and $CLS_3$. These results show that the SMNP model improves probabilistic prediction when asymmetric choice responses are present.

\begin{table}[tb!] 
\centering 
\caption{Predictive performance numerical experiments} \label{tab:MSNpredictive} 
\begin{tabular}{lrrrrrr} 
\toprule 
& \multicolumn{3}{c}{In-sample} & \multicolumn{3}{c}{Out-of-sample} \\ 
\cmidrule(lr){2-4} \cmidrule(lr){5-7} 
Metric & MNP & SMNP & Oracle & MNP & SMNP & Oracle \\ 
\midrule 
\multicolumn{7}{l}{\textit{Panel A: Asymmetric choice responses}} \\
\midrule 
$HR$ & $0.5470$ & $\mathbf{0.5483}$ & $0.5469$ & $0.5480$ & $\mathbf{0.5499}$ & $0.5492$ \\ 
$LS$ & $-0.9201\rlap{$^{*}$}$ & $\mathbf{-0.9166}$ & $-0.9166$ & $-0.9216\rlap{$^{*}$}$ & $\mathbf{-0.9192}$ & $-0.9192$ \\ 
$CLS_{1}$ & $-0.5707$ & $\mathbf{-0.5700}$ & $-0.5703$ & $-0.5704$ & $\mathbf{-0.5700}$ & $-0.5701$ \\ 
$CLS_{2}$ & $-0.5984$ & $\mathbf{-0.5978}$ & $-0.5977$ & $-0.5981$ & $\mathbf{-0.5973}$ & $-0.5974$ \\ 
$CLS_{3}$ & $-0.4725\rlap{$^{*}$}$ & $\mathbf{-0.4693}$ & $-0.4693$ & $-0.4750\rlap{$^{*}$}$ & $\mathbf{-0.4727}$ & $-0.4726$ \\
\midrule \multicolumn{7}{l}{\textit{Panel B: Symmetric choice responses}} \\ 
\midrule 
$HR$ & $\mathbf{0.6095}$ & $0.6078$ & $0.6083$ & $\mathbf{0.6071}$ & $0.6059$ & $0.6072$ \\ 
$LS$ & $-0.8824$ & $\mathbf{-0.8821}$ & $-0.8824$ & $\mathbf{-0.8837}$ & $-0.8838$ & $-0.8835$ \\ 
$CLS_{1}$ & $-0.4696$ & $\mathbf{-0.4696}$ & $-0.4696$ & $\mathbf{-0.4608}$ & $-0.4608$ & $-0.4607$ \\ 
$CLS_{2}$ & $-0.4621$ & $\mathbf{-0.4619}$ & $-0.4622$ & $\mathbf{-0.4724}$ & $-0.4724$ & $-0.4722$ \\ 
$CLS_{3}$ & $-0.6425$ & $\mathbf{-0.6423}$ & $-0.6425$ & $\mathbf{-0.6420}$ & $-0.6421$ & $-0.6421$ \\
\bottomrule 
\end{tabular}\\
\vspace{1ex} 
\raggedright
\footnotesize{\textit{Notes:} Bolded values indicate the superior predictive performance between the SMNP and MNP specifications within each DGP and sample split. The symbol $^{*}$ indicates that the difference between the model and the oracle is statistically significant at the $5\%$ significance level according to a \cite{giacomini2006tests} test on the differences in proper scoring rules.} 
\end{table}

\subsection{Results with symmetric choice responses}
\begin{figure}[tb!]
\centering
\includegraphics[width=\textwidth]{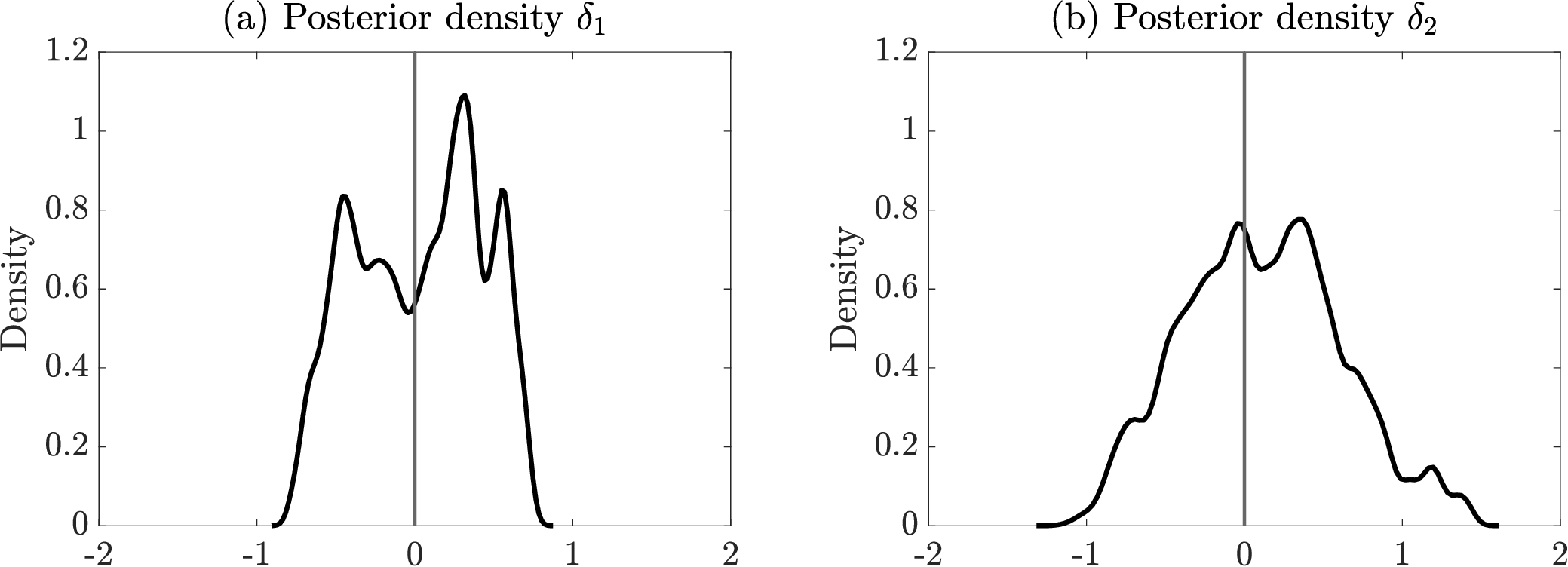}
\caption{Posterior densities $\delta_1$ and $\delta_2$ in the SMNP model for the experiment with symmetric latent utilities. The vertical lines represent true parameter values.}
\label{fig:idendeltaMVN}
\end{figure}

Second, we examine the experiment in which the data are generated from the MNP model. Figure~\ref{fig:idendeltaMVN} shows the posterior densities of the skewness parameters in the SMNP model. The posterior mass is centered around zero, indicating that the model recovers the symmetric latent utility structure. This also follows from the posterior choice probabilities reported in Appendix~\ref{A:numerical}, which are nearly identical between the MNP and SMNP models. In terms of predictive performance, Panel B of Table~\ref{tab:MSNpredictive} shows that the MNP model performs slightly better. However, the log scores and censored likelihood scores of both models are close to the oracle values, and the Giacomini--White tests do not reject equal predictive accuracy relative to the oracle.

Taken together, the two experiments show that the flexibility of the SMNP model is useful when asymmetry is present, but does not impose a meaningful predictive cost when the standard symmetry assumption is valid. When the latent utilities are skewed, it recovers the asymmetric choice responses and improves predictive accuracy relative to the MNP model. When the latent utilities are symmetric, it shrinks toward the MNP special case and does not sacrifice predictive performance. In sum, the SMNP model provides a compelling alternative to the MNP model: it helps avoid potential misspecification from symmetric latent utilities while allowing researchers to capture more realistic asymmetric choice responses.

\section{Empirical applications}\label{empirical}
This section demonstrates the practical relevance of our proposed SMNP model by applying it to two consumer choice datasets. Section~\ref{detergent} examines a laundry detergent purchase dataset and Section~\ref{ketchup} a ketchup purchase dataset. The implementation and prior settings are discussed in Section~\ref{sec:design}. We discuss posterior distributions estimated on the full data sets. 
For the evaluation of predictive performance, we repeatedly randomly  split the data into $80\%$ training and $20\%$ test sets. For each repetition $t=1,\dots,300$, the model is estimated on the training sample and we compute both in-sample and out-of-sample predictive metrics. We report average metrics across the repeated sampling iterations, and apply Giacomini--White tests to $\{HR^{(t)},LS^{(t)},CLS^{(t)}\}^{300}_{t=1}$ of SMNP and MNP.

\subsection{Laundry detergent purchases}\label{detergent}
The laundry detergent dataset includes $2,657$ observations of consumer purchases across six laundry detergent brands, together with the log price per ounce of each brand. The brands in the dataset are Tide $(26.38\%)$, EraPlus $(19.08\%)$, Surf $(15.28\%)$, Solo $(9.52\%)$, All $(3.27\%)$, and Wisk $(26.46\%)$. \citet{chintagunta1998empirical} discuss the data, and the data are available in \cite{imai2005mnp}. We follow previous analyses of this dataset and include brand intercepts and log prices in the latent utility specification \citep{imai2005bayesian,burgette2021symmetric,loaiza2022scalable,loaiza2023fast}.

\begin{figure}[tb!]
\centering
\includegraphics[width=\textwidth]{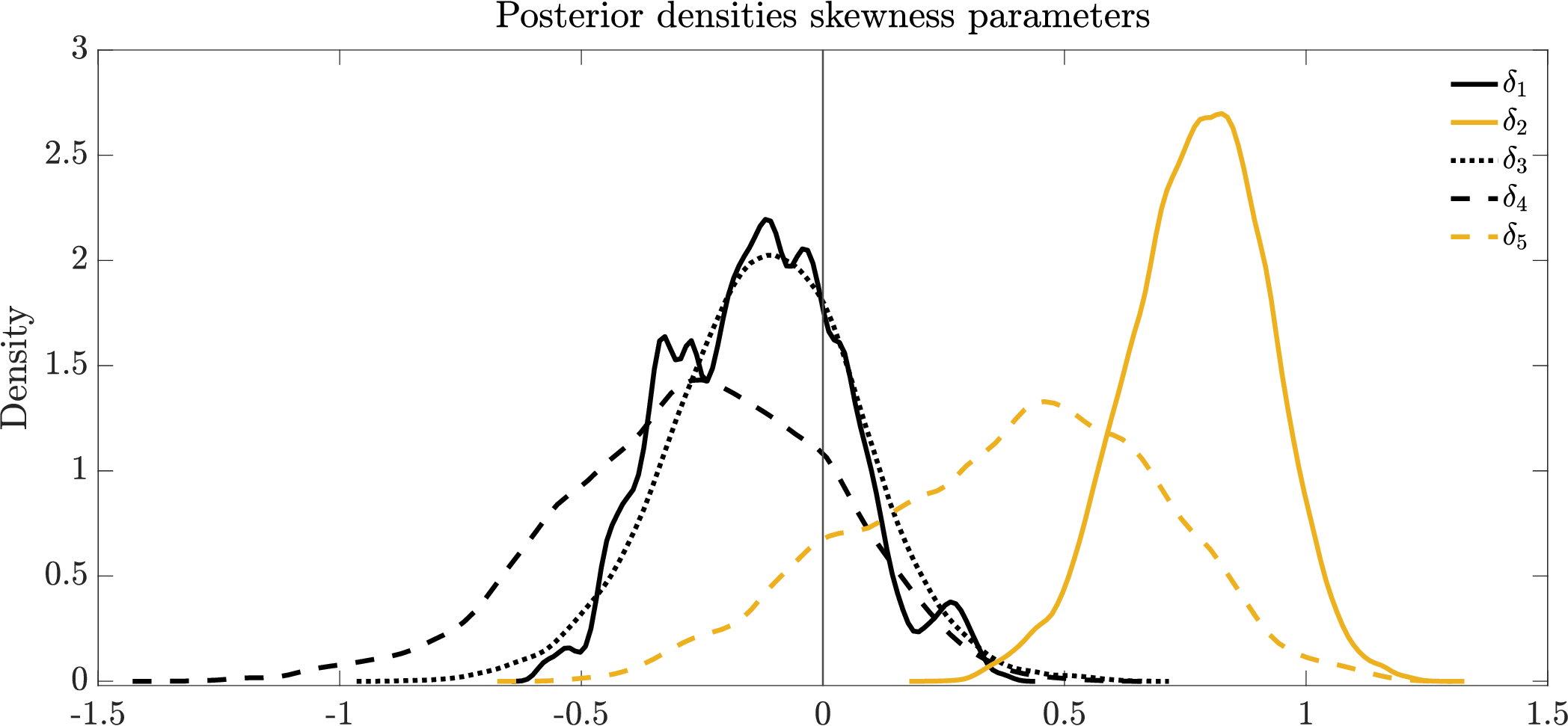}
\caption{Posterior densities for $\delta_1$ to $\delta_5$, representing the skewness parameters for detergent brands Tide, EraPlus, Surf, Solo, and All, respectively, with Wisk the base category. }
\label{fig:detergent_delta}
\end{figure}

The SMNP model picks up asymmetry in the latent utility distribution. Figure \ref{fig:detergent_delta} shows the posterior densities of the skewness parameters in the SMNP model. We find that the posterior density for $\delta_2$ allocates all probability mass at positive values, and  for $\delta_5$ most probability mass is also allocated to the right of zero. The posterior densities for $\delta_1, \delta_3$, and $\delta_4$ allocate high posterior density at zero. These results suggest asymmetry in the marginal latent utility distribution of EraPlus relative to the base category Wisk. 

\begin{figure}[tb!]
\centering
\includegraphics[width=\textwidth]{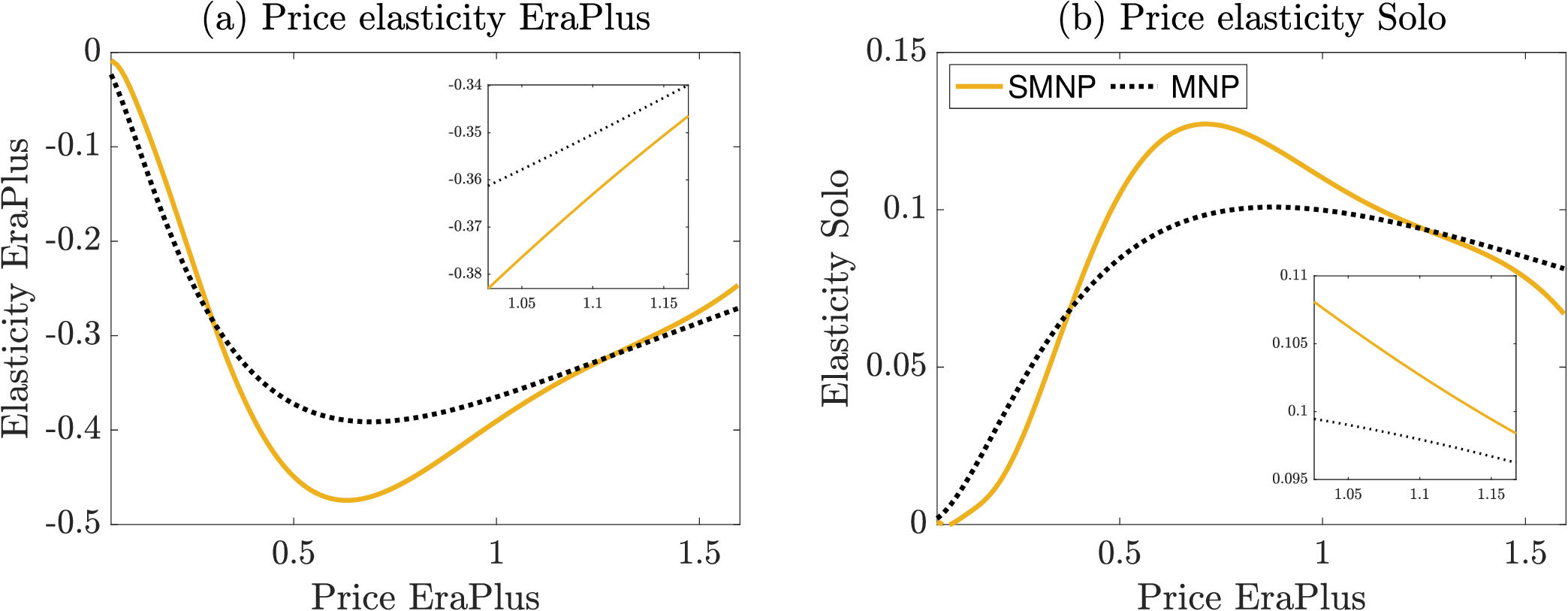}
\caption{Price elasticity of EraPlus and Solo given the price of EraPlus, with the prices of the other brands fixed at their mean. The solid yellow and dotted black lines represent the estimated elasticities using the SMNP and MNP models, respectively. The price elasticities are constructed from the choice probability functions, which are provided in Appendix~\ref{A:application}.}
\label{fig:elasticity_detergent}
\end{figure}

To investigate how this asymmetry impacts choice responses, we examine how the price of EraPlus affects choice probabilities. Figure~\ref{fig:elasticity_detergent} shows the price elasticities implied by the SMNP and MNP models as the price of EraPlus changes. Panel (a) reports the own-price elasticity for EraPlus and Panel (b) the cross-price elasticity for Solo.
The SMNP implies more nonlinear elasticity functions than the MNP model. For prices between 0.5 and 1, the SMNP model predicts a sharper decline in demand for EraPlus and a stronger shift toward Solo, indicating that the MNP model may understate both demand losses and competitive substitution. For lower or higher prices, the SMNP and MNP show similar price elasticities.

Pricing decisions are usually made around a specific current price. At the sample mean price of EraPlus ($1.06$), the SMNP implies stronger price sensitivity than the MNP model. The own-price elasticity of EraPlus is larger in magnitude, and the cross-price elasticity for Solo is also higher. Hence, a manager using the MNP around this price would understate both the demand loss from a price increase and the substitution toward Solo. The same logic applies in the opposite direction: for a price reduction, the MNP would understate the potential gain in demand for EraPlus. The SMNP model therefore provides a more informative basis for weighing the risks of price increases against the gains from discounts.

The predictive results in Table~\ref{tab:emppredictive} support this interpretation. For the laundry detergent dataset, the SMNP model improves the hit-rate, log score, and most censored likelihood scores both in-sample and out-of-sample. These improvements are mostly statistically significant. Hence, the differences between SMNP and MNP are not only economically meaningful, but they are also linked to better predictions of SMNP. In addition, we also apply the Giacomini--White test directly to the

\begin{table}[tb!]
\centering
\caption{Predictive performance empirical applications}
\label{tab:emppredictive}
\setlength{\tabcolsep}{3.5pt}
\begin{tabular}{lrrrrrrrr}
\toprule
& \multicolumn{4}{c}{Laundry detergent} & \multicolumn{4}{c}{Ketchup} \\
\cmidrule(lr){2-5} \cmidrule(lr){6-9}
& \multicolumn{2}{c}{In-sample} & \multicolumn{2}{c}{Out-of-sample}
& \multicolumn{2}{c}{In-sample} & \multicolumn{2}{c}{Out-of-sample} \\
\cmidrule(lr){2-3} \cmidrule(lr){4-5} \cmidrule(lr){6-7} \cmidrule(lr){8-9}
Metric & MNP & SMNP & MNP & SMNP & MNP & SMNP & MNP & SMNP \\
\midrule
$HR$      & $0.4963$ & $\mathbf{0.4974}$ & $0.4958$ & $\mathbf{0.4967}$ & $0.5805$ & $\mathbf{0.5806}$ & $0.5802$ & $\mathbf{0.5804}$ \\
$LS$      & $-1.3183$ & $\mathbf{-1.3153}\rlap{$^{*}$}$ & $-1.3231$ & $\mathbf{-1.3202}\rlap{$^{*}$}$ &  $-0.9813$ & $\mathbf{-0.9803}\rlap{$^{*}$}$ & $-0.9832$ & $\mathbf{-0.9826}\rlap{$^{*}$}$ \\
$CLS_{1}$ & $-0.4678$ & $\mathbf{-0.4677}\rlap{$^{*}$}$ & $-0.4689$ & $\mathbf{-0.4689}$ & $-0.4323$ & $\mathbf{-0.4321}\rlap{$^{*}$}$ & $\mathbf{-0.4328}$ & $-0.4328$ \\
$CLS_{2}$ & $-0.3966$ & $\mathbf{-0.3940}\rlap{$^{*}$}$ & $-0.3967$ & $\mathbf{-0.3942}\rlap{$^{*}$}$ & $-0.1729$ &  $\mathbf{-0.1719}\rlap{$^{*}$}$ & $-0.1741$  & $\mathbf{-0.1732}\rlap{$^{*}$}$ \\
$CLS_{3}$ & $-0.3185$ & $\mathbf{-0.3174}\rlap{$^{*}$}$ & $-0.3191$ & $\mathbf{-0.3180}\rlap{$^{*}$}$ & $\mathbf{-0.4778}\rlap{$^{*}$}$ & $-0.4778$ & $\mathbf{-0.4782}\rlap{$^{*}$}$ & $-0.4783$  \\
$CLS_{4}$ & $-0.2823$ & $\mathbf{-0.2815}\rlap{$^{*}$}$ & $-0.2832$ & $\mathbf{-0.2825}\rlap{$^{*}$}$ & $-0.5853$ & $\mathbf{-0.5852}\rlap{$^{*}$}$ & $-0.5858$ & $\mathbf{-0.5858}\rlap{$^{*}$}$ \\
$CLS_{5}$ & $-0.1273$ & $\mathbf{-0.1271}\rlap{$^{*}$}$ & $-0.1291$ & $\mathbf{-0.1289}\rlap{$^{*}$}$ & -- & -- & -- & -- \\
$CLS_{6}$ & $-0.4733$ & $\mathbf{-0.4728}\rlap{$^{*}$}$ & $-0.4744$ & $\mathbf{-0.4738}\rlap{$^{*}$}$ & -- & -- & -- & -- \\
\bottomrule 
\end{tabular}\\ \vspace{1ex} \raggedright 
\footnotesize{ \textit{Notes:} The predictive performance measures are averaged across $T=300$ repeated random sample splitting iterations. For laundry detergent, $CLS_1$ to $CLS_6$ refer to Tide, EraPlus, Surf, Solo, All, and Wisk. For ketchup, $CLS_1$ to $CLS_4$ refer to Hunts, Del Monte, STB, and Heinz. Bolded values indicate superior predictive performance between the MNP and SMNP specifications within each application and sample split. The symbol $^{*}$ denotes a statistically significant difference in predictive performance.} \end{table}

\subsection{Ketchup purchases}\label{ketchup}
The ketchup dataset consists of $4,956$ observations of consumer purchases across four ketchup brands. The ketchup brands are Hunts $(20.56\%)$, Del Monte $(5.17\%)$, a store brand referred to as STB $(23.31\%)$, and Heinz $(50.97\%)$. \citet{kim1995modeling} discuss the data, and the data is available in \cite{ecdat}. We maintain the same utility specification with brand intercepts and log prices. 

\begin{figure}[tb!]
\centering
\includegraphics{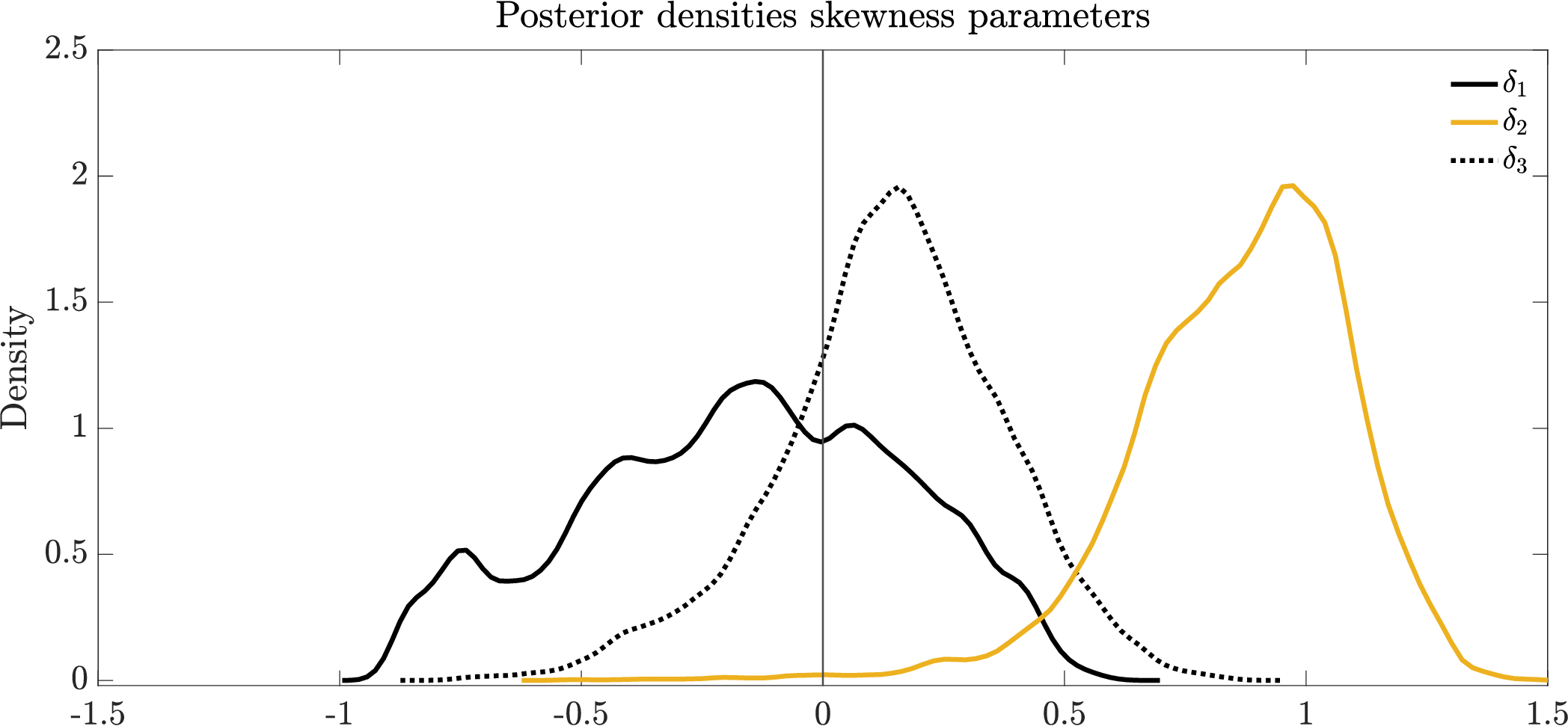}
\caption{Posterior densities for $\delta_1$ to $\delta_3$, representing the skewness parameters for ketchup brands Hunts, Del Monte, and STB respectively, with Heinz the base category.}
\label{fig:ketchup_delta}
\end{figure}

This application provides a second test of whether asymmetric latent utilities matter in consumer choice data. Figure~\ref{fig:ketchup_delta} shows that the posterior density of $\delta_2$, corresponding to Del Monte relative to the base category Heinz, places most of its mass above zero. The posteriors for Hunts and the store brand are located closer to zero. The strongest evidence of asymmetry therefore again appears for a specific alternative rather than uniformly across all brands.

\begin{figure}[tb!]
\centering
\includegraphics[width=\textwidth]{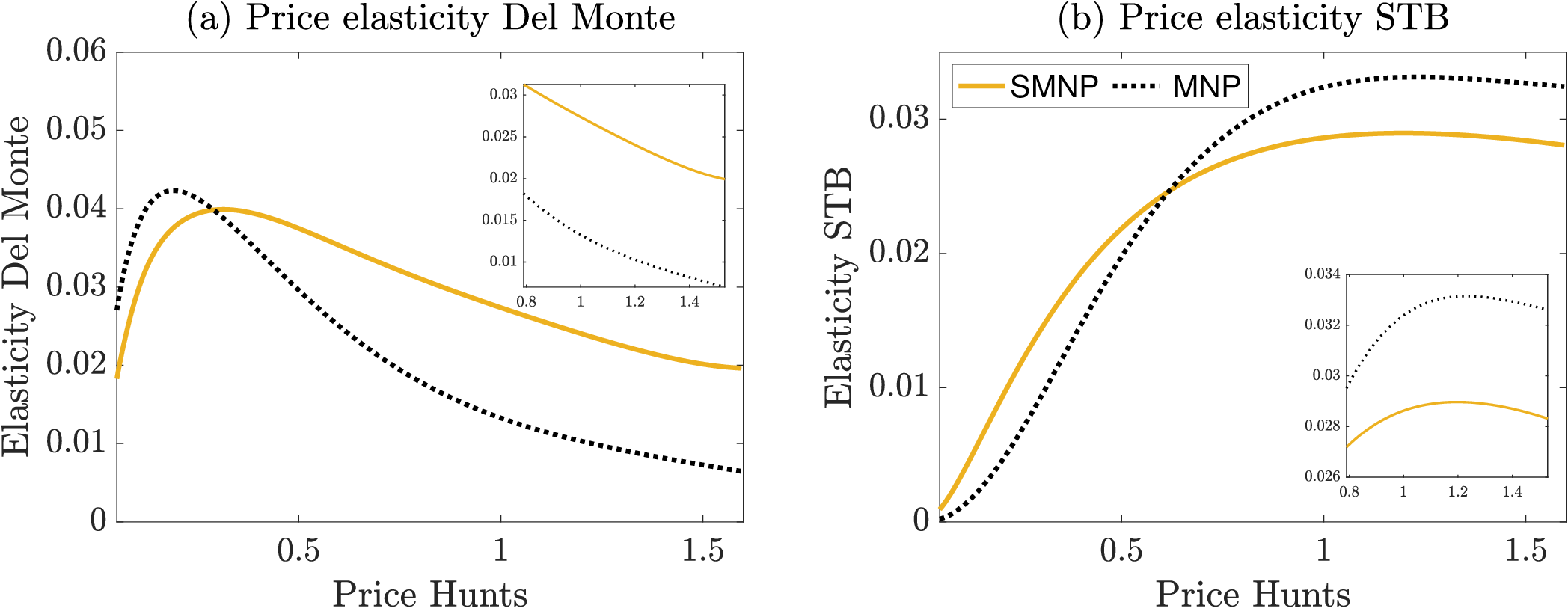}
\caption{Price elasticity given the price of Hunts indicated on the x-axis.  The solid yellow and dotted black lines represent the estimated elasticity using the SMNP and MNP models, respectively.}
\label{fig:elasticity_ketchup}
\end{figure}

Figure~\ref{fig:elasticity_ketchup} shows how the elasticities differ between the SMNP and MNP models. Panel (a) shows the cross-price elasticity for Del Monte and Panel (b) for STB, as a function of the price for Hunts. The SMNP changes which competing brands are predicted to benefit from a price change for Hunts. Around the sample mean price of Hunts ($1.34$), it implies a higher cross-price response for Del Monte and a lower cross-price response for STB relative to the MNP model. Hence, the MNP may not only misstate the magnitude of price sensitivity, but it may also misallocate the predicted substitution pattern across competing brands.

The predictive results in Table~\ref{tab:emppredictive} also favor the SMNP model. The SMNP improves the  hit rates and log scores both in-sample and out-of-sample.  The largest predictive gain occurs for $CLS_2$, corresponding to Del Monte, which aligns with the posterior evidence of nonzero skewness for $\delta_2$. The MNP model has a higher censored likelihood score for the third category corresponding to the store brand, which has a posterior skewness parameter density centered close to zero. This suggests that the model is most useful for alternatives where the data contain evidence of asymmetric choice responses.

Taken together, the two applications suggest that the SMNP model can help practitioners avoid misleading price-sensitivity estimates and make better pricing and substitution decisions than those based on the MNP model alone. The detergent application shows how asymmetric latent utilities can change the estimated shape and magnitude of own- and cross-price elasticities for a focal brand. The ketchup application shows that the same modeling flexibility can alter the implied allocation of substitution across competitors. In both cases, the SMNP model improves prediction.

\section{Discussion} \label{discuss}
This paper extends the MNP by allowing the conditional distribution of the latent utilities to be multivariate skew-normal, in which the degree and direction of asymmetry can differ across alternatives. This preserves the main advantage of the MNP: Latent utilities can be correlated and it therefore captures substitution patterns across alternatives. At the same time, it relaxes the symmetry restriction imposed by the multivariate normal distribution. 

The main methodological contribution is to make this extension identified and estimation feasible. We propose a reparameterization that enforces the scale normalization and positive definiteness of the covariance matrix. We also develop a Bayesian estimation strategy based on double data augmentation. The resulting MCMC sampler only includes a single one-dimensional Metropolis-Hastings step, and updates all other parameters using Gibbs.

Numerical experiments and empirical applications demonstrate that the SMNP model is a useful tool for multinomial choice analysis. It nests the standard MNP model, captures asymmetric choice responses, and provides economically meaningful differences in elasticities and substitution patterns. The model offers a parsimonious way to reduce misspecification risk and obtain more reliable predictions for pricing and policy decisions.

The SMNP model captures asymmetric choice responses through the distribution of latent utilities, but it does not by itself identify the behavioral mechanism generating the asymmetry. The empirical evidence that asymmetric choice responses matter creates opportunities for future research. For instance, approximate Bayesian methods may be useful for larger choice sets or richer SMNP models with random coefficients. In these settings, the proposed exact Bayesian sampler may suffer from slow mixing due to the double data augmentation and the Metropolis-Hastings update for the scale-restricted skewness parameter.

\bibliography{cash.bib}
\clearpage
   
\appendix  
\section*{Appendix}
    
\section{Prior elicitation for \texorpdfstring{$\delta$}{delta} and \texorpdfstring{$\Sigma$}{Sigma}}\label{prioreliccit}

\subsection{Derivation of the identified covariance parameterization}
This appendix derives the covariance parameterization in Section~\ref{identification}. Starting from a Cholesky decomposition of $\Sigma_u$, we show how the scale normalization can be imposed directly in the augmented parameter space. The derivation motivates placing priors on $\{\delta,\gamma,\psi\}$, because draws of these parameters automatically produce a positive definite $\Sigma_u$ and an implied scale matrix $\Sigma=\Sigma_u+\delta\delta^\top$ satisfying $\Sigma_{11}=1$.

Since $\Sigma_u$ is a covariance matrix, it is symmetric and positive definite, and therefore admits a unique decomposition $\Sigma_u = LL^\top$, where $L$ is a lower triangular matrix with real and positive diagonal entries. We define
\begin{equation}
\begin{aligned}
 L
&=\begin{bmatrix}
L_{11} & 0\\
L_{21} &  L_{22} 
\end{bmatrix},
\end{aligned}
\end{equation}
where $L_{11}$ is a positive scalar, $L_{21}$ is a $(J-1) \times 1$ column vector, $L_{22}$ is a $(J-1) \times (J-1)$ lower-triangular matrix with real and positive diagonal entries. Thus, we have 
\begin{equation}
\begin{aligned}
\Sigma_u 
=\begin{bmatrix}
L_{11} & 0\\
L_{21} &  L_{22} 
\end{bmatrix}
\begin{bmatrix}
L_{11} & L_{21}^\top \\
0 &  L_{22} ^\top 
\end{bmatrix}
=\begin{bmatrix}
L_{11}^2 & L_{11}L_{21}^\top \\
L_{21}L_{11} &  L_{21}L_{21}^\top +L_{22}L_{22}^\top 
\end{bmatrix}.
\end{aligned}
\end{equation}
To enforce scale identification, we impose the restriction $L_{11}^2=1-\delta^2_1$, which implies $L_{11}=\sqrt{1-\delta^2_1}$. Let $L_{21}L_{11}=\gamma$, so that $L_{21}=\frac{\gamma}{\sqrt{1-\delta^2_1}}$. Since $L_{22}$ is a lower-triangular matrix with real and positive diagonal entries, we have $L_{22}L_{22}^\top= \psi$ is a positive-definite matrix. Under this parameterization, $\Sigma_u$ can be expressed as 
\begin{equation}
\Sigma_u=
\begin{bmatrix}
1-\delta^2_1 & \gamma^\top\\
\gamma &  \psi+ \frac{\gamma\gamma^\top}{1-\delta^2_1} 
\end{bmatrix}.
\end{equation} 

This parameterization allows us to place priors directly on $(\delta, \gamma, \psi)$. Sampling these parameters iteratively in the MCMC enables us to construct posterior draws of $\Sigma_u$ that automatically satisfy the identification restriction. For the symmetric matrix $\Sigma_u$, necessary and sufficient conditions for positive definiteness can be obtained via the Schur complement theorem \citep{boyd2004convex}. Specifically, $\Sigma_u \succ 0$ if and only if (i) $1-\delta_1^2 > 0$ and (ii) the associated Schur complement $\psi + \frac{\gamma \gamma^\top}{1-\delta_1^2} - \frac{\gamma \gamma^\top}{1-\delta_1^2} = \psi \succ 0$, where the notation $ \mathcal{Q} \succ 0$ indicates that the matrix $\mathcal{Q}$ is positive definite. The first condition implies a linear constraint on the first skewness parameter such that $\delta_1 \in (-1,1)$. 

\subsection{Constructing priors for \texorpdfstring{$\delta$}{delta} and \texorpdfstring{$\Sigma$}{Sigma}}
The prior hyperparameters are set such that the implied prior mean of $\Sigma$ is the equicorrelated matrix $\mathcal{M}=\frac{1}{2} 1_J1_J^\top + \frac{1}{2}I_J$. Since $\Sigma = \Sigma_u + \delta \delta^\top$, we impose
\begin{align}  
\mathbb{E}[\Sigma_u]
=\mathcal{M}-\mathbb{E}[\delta\delta^\top] = \mathcal{M}-\mathbb{V}[\delta],
\end{align} 
where we use that the prior mean of $\delta$ is zero. We can write
\begin{equation} \label{eq:Vdelta}    
\mathbb{V}[\delta]=
\begin{bmatrix}
\mathbb{V}(\delta_1)& 0_{J-1}^\top\\
0_{J-1}&  \tau_\delta I_{J-1}
\end{bmatrix},
\end{equation} 
and $ \mathbb{V}(\delta_1)=\tau_\delta\left[1-\frac{\frac{1}{\sqrt{\tau_\delta}}\varphi\!\big(\frac{1}{\sqrt{\tau_\delta}}\big)}{\Phi\!\big(\frac{1}{\sqrt{\tau_\delta}}\big)-\tfrac12}\right]$ follows from the assumed prior on $\delta$. Here, $\varphi(\cdot)$ is the probability density function of a standard normal distribution.
Furthermore, $\mathcal{M}$ can be partitioned as 
\begin{equation}  
\begin{aligned}
\mathcal{M}
&=
\begin{bmatrix}
1 & 0.5\times1_{J-1}^\top\\
0.5\times1_{J-1} &  0.5 \times1_{J-1}1_{J-1}^\top + 0.5I_{J-1}
\end{bmatrix},
\end{aligned}
\end{equation}
which implies the following constraint on the prior of $\Sigma_u$: 
\begin{equation}\label{eq:exp_sig_u1}
\begin{aligned}
\mathbb{E}[\Sigma_u]
&=
\begin{bmatrix}
1- \mathbb{V}(\delta_1)& 0.5\times 1_{J-1}^\top\\
0.5 \times 1_{J-1}& 0.5\times 1_{J-1}1_{J-1}^\top + (0.5- \tau_\delta)I_{J-1} 
\end{bmatrix}.\\
\end{aligned}
\end{equation}
Due to the parametrization of $\Sigma_u$ in \eqref{eq:sig_u}, the constraint in \eqref{eq:exp_sig_u1} implies three separate constraints. First, $\mathbb{E}(1-\delta_1^2) = 1-\mathbb{V}(\delta_1)$, which always follows by the fact that $\mathbb{E}(\delta_1^2)=\mathbb{V}(\delta_1)$. Second $\mathbb{E}(\gamma) = 0.5\times 1_{J-1}$, which can be easily imposed by setting $B_\gamma = 0.5\times 1_{J-1}$.
Third, we must impose the condition 
\begin{equation} \label{eq:cond3}
\begin{split}  
\mathbb{E}\left[\psi+ \frac{\gamma\gamma^\top}{1-\delta^2_1}\right]
&= 0.5\times 1_{J-1}1_{J-1}^\top + (0.5-\tau_\delta)I_{J-1},\\
\end{split}
\end{equation} 
where $\tau_\delta$ must be chosen such that the right-hand side of \eqref{eq:cond3} is positive definite. It is straightforward to show that the smallest eigenvalue of this matrix is $0.5-\tau_\delta$, and hence $\tau_\delta<0.5$ guarantees positive definiteness.

Next, since we assume independence between $\gamma$ and $\delta$ apriori, \eqref{eq:cond3} can be written as 
\begin{equation}\label{Eq22block} 
\begin{split}   
\mathbb{E}\left[\psi+ \frac{\gamma\gamma^\top}{1-\delta^2_1}\right] 
&=  \frac{V}{(J+3)-(J-1)-1}+ \mathbb{E}\left[\frac{1}{1-\delta^2_1}\right] \left[\tau_\gamma I_{J-1} + 0.5^21_{J-1}1_{J-1}^\top
\right],\\
\end{split}
\end{equation} 
Next, replace \eqref{Eq22block} in \eqref{eq:cond3} and solve for $V$ to get.
\begin{equation} \label{eq:V2}
\begin{split}   
V
&= 3\left[\left(0.5 -  \mathbb{E}\left[\frac{1}{1-\delta^2_1}\right] 0.5^2  \right)1_{J-1}1_{J-1}^\top + (0.5-\tau_\delta -  \mathbb{E}\left[\frac{1}{1-\delta^2_1}\right] \tau_\gamma)I_{J-1} \right].
\end{split}
\end{equation} 
The matrix $V$ is positive definite if and only if 
\begin{align}
    \left[\left(0.5 -  \mathbb{E}\left[\frac{1}{1-\delta^2_1}\right] 0.5^2  \right)1_{J-1}1_{J-1}^\top + (0.5-\tau_\delta -  \mathbb{E}\left[\frac{1}{1-\delta^2_1}\right] \tau_\gamma)I_{J-1} \right] \succ 0.
\end{align} 
The upper bound of $\tau_\gamma$ that guarantees the eigenvalues of $V$ to be positive is
\begin{equation} \label{eq:UB_taugamma}
\begin{split}  
\text{UB}_\gamma< \text{min}\left(\frac{0.5+ q\times0.5-\tau_\delta- q  \mathbb{E}\left[\frac{1}{1-\delta^2_1}\right] 0.5^2 }{ \mathbb{E}\left[\frac{1}{1-\delta^2_1}\right]},\frac{0.5-\tau_\delta}{ \mathbb{E}\left[\frac{1}{1-\delta^2_1}\right]} \right),
\end{split}
\end{equation} 
where the expectation $\mathbb{E}\left[\frac{1}{1-\delta^2_1}\right]  = \int^{c}_{-c} \frac{1}{1-\delta^2_1} f(\delta_1)d\delta_1 $ 
with  $f(\delta_1) =  \frac{\phi_1\left(\frac{\delta_1}{\sqrt{\tau_\delta}}\right)}{\sqrt{\tau_\delta}\,\left[2\Phi\left(\tfrac{1}{\sqrt{\tau_\delta}}\right)-1\right]} $ can be easily computed numerically. In practice we set $\tau_\gamma=0.9\text{UB}_\gamma$ for computational stability.

\subsection{Calibrating \texorpdfstring{$\tau_\delta$}{tau-delta}}
Recall $\Sigma = \Sigma_u + \delta\delta^\top$, which can be written explicitly as 
\begin{equation} \label{eq:Sigma2}
\begin{aligned}
\Sigma
&=
\begin{bmatrix}
1 & \gamma^\top + \delta_1\delta_{2:J}^\top\\
\gamma +  \delta_1\delta_{2:J} &  \psi+ \frac{\gamma\gamma^\top}{1-\delta^2_1 } + \delta_{2:J}\delta_{2:J}^\top 
\end{bmatrix}.
\end{aligned}
\end{equation}
The hyperparameter $\tau_\delta$ governs the degree of asymmetry allowed by the prior. As $\tau_\delta$ approaches $0$, the prior on $\delta$ becomes more concentrated around zero. In this limiting case, the prior specification reduces the model to the MNP of \citep{mcculloch2000bayesian}, since $\delta=0$ in \eqref{eq:Sigma2} yields the MNP covariance matrix structure. Conversely, as $\tau_\delta$ approaches its theoretical upper bound of $0.5$, the upper bound on $\tau_\gamma$ in \eqref{eq:UB_taugamma} collapses to zero. This implies that the MNP cannot be recovered, as some latent utility correlations can only be captured via the asymmetry parameters $\delta$, which necessarily induces deviations from normality.

We select a value of $\tau_\delta$ that does not favor either specification. Specifically, we set $\tau_\delta = 0.1936$, which closely approximates $\text{UB}_{\gamma}$ and ensures that both $\tau_\gamma$ and $\tau_\delta$ have non-negligible magnitude.

\section{MCMC sampling scheme}\label{MCMC}
\subsection{Initialization}\label{initial}
This section outlines the initial values for the model parameters
$\{\delta_{1:J}^{(0)},w^{(0)},\gamma^{(0)},\psi^{(0)},Z^{(0)},\beta^{(0)}\}$. We initialize $\delta$ and $w$ as $J-$dimensional and $N-$dimensional zero vectors, $0_J$ and $0_N$, respectively. We set the initial value for $\Sigma$ to be a $J$-dimensional identity matrix, implying that $\Sigma_u$ is also an identity matrix. As a result, the initial values for $\psi$ and $\gamma$ can be calculated using Equation \ref{eq:sig_u}.

Next, we initialize the differenced utilities. We begin by drawing the undifference utilities independently from a standard normal distribution, such that $\bar{z}_{ik}\sim N(0,1)$ for all alternatives $k=1,\dots, J+1$. We assign the maximum simulated value to the chosen alternative $Y_i$. Finally, the starting values for the differenced utilities are calculated by subtracting the baseline alternative $J+1$: $z_{ij} = \bar{z}_{ij}-\bar{z}_{iJ+1}$  for $j = 1, \dots, J$.

We initialize the alternative-specific intercepts in $\beta$ using the observed choice frequencies. Let $n_j=\sum_{i=1}^N \mathbb{I}(Y_i=j)$ denote the number of observations choosing alternative $j=1,\ldots,J+1$. For each non-base alternative, we choose the initial intercept that matches the empirical frequency in a binary probit comparison between alternative $j$ and the base alternative: $\beta_j^{(0)}=\Phi^{-1}\left(\frac{n_j}{n_j+n_{J+1}}\right)$. We initialize the remaining slope coefficients at zero.

\subsection{Details on the sampling steps}\label{sampler}
We elucidate the steps of the sampling scheme in this appendix. The objective is to sample the model parameters $\{\beta,\delta,\psi,\gamma\}$, as well as the latent variables $\{Z,w\}$. The joint posterior distribution can be written as $p(\beta, \psi,\gamma,\delta,Z,w \mid Y,X)
\propto$
\begin{equation}\label{eq:post2}
 \prod^N_{i=1} p(Y_i\mid Z_i) p(Z_i,w_i\mid X_i,\beta,\psi, \delta, w_i>0)p(\beta)p(\psi)p(\gamma)p({\delta}) I\left(\left|\delta_1 \right|< c\right),\\
\end{equation}
which is a reparametrized version of the posterior in \eqref{eq:augpost}. Since $\Sigma$ can be uniquely determined by $\{\delta,\psi,\gamma\}$, sampling these parameters is equivalent to sampling $\Sigma$. In our MCMC, we iterate over the following steps.
 
\begin{description}
    \item[\textbf{Step 1:}] Generate $\delta_1^{(r)} \sim p(\delta_1 \mid \delta_{2:J}^{(r-1)}, w^{(r-1)},\gamma^{(r-1)},\psi^{(r-1)}, Z^{(r-1)}, \beta^{(r-1)},Y,X)$.
\end{description}
First, we ensure the constraint $\left|\delta_1 \right|< c$ is adhered by introducing the transformation
\begin{equation}\label{eq:function}
\begin{split}
\delta_1 = g(\tilde{\delta}_1)=c\tanh(\tilde{\delta}_1),
\end{split}
\end{equation}
where $g:\mathbb{R} \to (-c,c)$ so $\tilde{\delta}_1 \in (-\infty,\infty)$. Hence, sampling is performed in an unconstrained space and the parameter of interest $\delta_1$ can be recovered via  \eqref{eq:function}. This step is a univariate MH step. 

Second, we generate a candidate draw $\tilde{\delta}_1^c$ from the proposal density $q(\tilde{\delta}_1)$. We then evaluate the acceptance probability 
\begin{equation}
\begin{split}
acc=min\left\{1,\frac{p(\tilde{\delta}_1^c\mid \beta , \psi,\gamma, Z,w,X)q(\tilde{\delta}_1^{(r-1)})}
{p(\tilde{\delta}_1^{(r-1)}\mid \beta , \psi,\gamma, Z,w,X)q(\tilde{\delta}_1^c)}\right\}.
\end{split}
\end{equation}
Next, we sample $U\sim Unif(0,1)$. If $U < acc$, we accept the candidate draw by setting $\tilde{\delta}_1^{(r)}=\tilde{\delta}_1^{c}$. Otherwise, the draw is rejected and we retain the previous value, $\tilde{\delta}_1^{(r)}=\tilde{\delta}_1^{(r-1)}$.

Here, the proposal density $q(\tilde{\delta}_1)$ is constructed via a Laplace approximation to the conditional posterior of $\tilde{\delta}_1$. Specifically, we numerically maximize the log of posterior \eqref{eq:condtildelta1} to obtain the Maximum A Posteriori (MAP) estimate $\tilde {\delta_1}_{MAP}$, which corresponds to the mode of the log posterior. Next, we compute the Hessian $H$ of the log posterior evaluated at $\tilde {\delta_1}_{MAP}$. A second-order Taylor expansion of the log posterior around $\tilde {\delta_1}_{MAP}$, followed by exponentiation, implies a normal kernel with mean $\tilde {\delta_1}_{MAP}$ and variance $-H^{-1}$, which is commonly used as a proposal density in a Metropolis-Hastings step. However, to allow for heavier tails in the proposal, we employ a Student-$t$ distribution with degree of freedom $5$, location $\tilde {\delta_1}_{MAP}$ and scale chosen such that the resulting variance matches $-H^{-1}$. The proposal density is 
\begin{equation}
\begin{split} 
q(\tilde{\delta}_1) = t_5\left( \tilde {\delta_1}_{MAP}, -\frac{3H^{-1}}{5} \right).
\end{split}
\end{equation}
The full conditional posterior of $\tilde{\delta}_1$ is given by 
\begin{equation}\label{eq:condtildelta1}
\begin{split}
p(\tilde{\delta}_1\mid \beta , \psi,\gamma, Z,w,X)
&\propto \prod^N_{i=1} p(Z_i(\tilde{\delta}_1)\mid X_i,\beta,\psi, \delta(\tilde{\delta}_1), w_i)p(\delta(\tilde{\delta}_1)) \left| \mathcal{J}(\tilde{\delta}_1) \right|  \\
&\propto \left| \Sigma_u(\tilde{\delta}_1) \right| ^{-\frac{n}{2}} \exp \left(-\frac{1}{2} \sum_{i=1}^N u_i(\tilde{\delta}_1)^\top\Sigma_u(\tilde{\delta}_1)^{-1}u_i(\tilde{\delta}_1)
\right)\times\\
&\ \ \ \ \  \exp\left\{-\frac{1}{2}\left(\delta(\tilde{\delta}_1)-B_\delta\right)^\top A_\delta\left(\delta(\tilde{\delta}_1)-B_\delta\right)
\right\} \left| \mathcal{J}(\tilde{\delta}_1) \right|,  
\end{split}
\end{equation}
with Jacobian transformation
\begin{equation}
\begin{split} 
\left| \mathcal{J}(\tilde{\delta}_1) \right| = \left|\frac{\partial \delta_1}{\partial\tilde{\delta}_1}\right|=c\left(1-tanh^2(\tilde{\delta}_1)\right).
\end{split}
\end{equation}

 \begin{description}
   \item[\textbf{Step 2:}] Generate $\delta_{2:J}^{(r)} \sim p(\delta_{2:J} \mid \delta_{1}^{(r)}, w^{(r-1)},\gamma^{(r-1)},\psi^{(r-1)}, Z^{(r-1)}, \beta^{(r-1)},Y,X)$.
\end{description} 
Let $\Sigma_u^{-1}=G = CC^\top$ via the Cholesky decomposition, and write 
\begin{equation} 
\begin{aligned} 
&C^\top Z_i = C^\top X_i\beta+C^\top w_i\delta  + C^\top u_i\\
&Z_i^*=X_i^*\beta+w_i^*\delta +u_i ^*, \quad \quad \quad u_i^*\sim N_J(0,I_J).
\end{aligned}
\end{equation}
The full conditional posterior of $\delta$ is given by:
\begin{equation}
\delta\mid \beta , \psi,\gamma, Z,w,X \sim N_J(\hat\delta,\Sigma_\delta)I\left(\left|\delta_1 \right|< c\right), \Sigma_\delta=(w^{*\top}w^*+A_\delta)^{-1}, \hat\delta=\Sigma_\delta(w^{*\top}Z^*_\delta+A_\delta B_\delta),
\end{equation}
where $Z^*_\delta=\left(\left(Z_1^*-X_1^*\beta)^\top,\dots,(Z_N^*-X_N^*\beta\right)^\top\right)^\top$. The full joint posterior distribution $p(\delta \mid \beta , \psi,\gamma, Z,w,X )$, can be expressed as the product of marginal and conditional densities:
\begin{equation}  
p(\delta \mid  \beta , \psi,\gamma, Z,w,X )=p(\delta_1\mid  \beta , \psi,\gamma, Z,w,X )p(\delta_{2:J}\mid \delta_1, \beta , \psi,\gamma, Z,w,X)I\left(\left|\delta_1 \right|< c\right),
\end{equation}
where $\delta_1$ is a scalar and $\delta_{2:J}$ is a $(J-1) \times 1$ vector. Let 
\begin{equation}  
\hat\delta=\begin{bmatrix}
\hat\delta_{_1} \\
\hat\delta_{_{2:J}}
\end{bmatrix},\quad\quad
\Sigma_\delta=\begin{bmatrix}
\Sigma_{\delta_1} &  \Sigma_{\delta_{(-1)(1)}}\\
\Sigma_{\delta_{(1)(-1)}} & \Sigma_{\delta_{(-1)(-1)}}
\end{bmatrix}.
\end{equation}
By the conditional property of the multivariate normal distribution, the conditional posterior of $\delta_{2:J}$ has a closed-form:
$$
\delta_{2:J}\mid \delta_1, \beta , \psi,\gamma, Z,w,X \sim N_{J-1}(\hat\delta_{{2:J}\mid \delta_1},\Sigma_{\delta_{2:J}\mid \delta_1})
$$
where $\hat\delta_{\delta_{2:J}\mid \delta_1}=\hat\delta_{\delta_{2:J}}+ \Sigma_{\delta_{(1)(-1)}}\Sigma_{\delta_1}^{-1}(\delta_1-\hat\delta_{\delta_1})$ and $\Sigma_{\delta_{2:J}\mid \delta_1}=\Sigma_{\delta_{(-1)(-1)}}- \Sigma_{\delta_{(1)(-1)}}\Sigma_{\delta_1}^{-1} \Sigma_{\delta_{(-1)(1)}}$.

\begin{description}
    \item[\textbf{Step 3:}] Generate $w^{(r)} \sim p(w \mid \delta^{(r)},\gamma^{(r-1)},\psi^{(r-1)}, Z^{(r-1)},\beta^{(r-1)},Y,X)$.
\end{description}
We follow \cite{fruhwirth2010bayesian} and sample $w_i$ independently by considering the linear model:
\begin{equation} 
\begin{aligned}
Z_i &= X_i\beta+\delta w_i + u_i, \\
u_i &\sim N_J(0, \Sigma_u), \quad w_i \sim N^+_0(0, 1).
\end{aligned}
\end{equation}
Here $N^+_{b}$ and $N^-_{b}$ represent a univariate normal distribution truncated from below or above by a bound $b$, respectively. We derive the full conditional posterior as follows:
\begin{equation}
\begin{split}
p(w_i \mid \beta,\delta,G, Z_i,X_i)
&\propto  p(Z_i\mid X_i\beta,G, \delta, w_i,X_i) p( w_i\mid  w_i>0)\\
&\propto \exp\left\{-\frac{1}{2}\left(Z_i-X_i\beta-\delta w_i\right)^\top\Sigma_u^{-1}\left(Z_i-X_i\beta-\delta w_i\right)
\right\}\times
\\&\phantom{\propto} \,\,\exp\left\{-\frac{w^2_i}{2}
\right\} \mathbb{I}(w_i>0).
\end{split}
\end{equation}

By completing the square, we obtain the conditional posterior of $w_i$ such that 
\begin{equation}
w_i \mid \beta,\delta,G, Z_i ,X_i\sim N_0^+(\hat{w},A_w),\quad A_w=(1+\delta^\top G\delta)^{-1}\quad \hat{w}= A_w\delta^\top G(Z_i-X_i\beta).
\end{equation}

\begin{description}
    \item[\textbf{Step 4:}] Generate $\gamma^{(r)} \sim p(\gamma \mid \delta^{(r)}, w^{(r)},\psi^{(r-1)}, Z^{(r-1)},\beta^{(r-1)},Y,X)$.
\end{description}    
In this step, we follow \cite{mcculloch2000bayesian} by first partitioning the error vector $u_i = \left(e_i,\eta_i  \right)^\top$, where $e_i$ is the first element of $u_i$ and $\eta_i$ corresponds to the remaining elements. We obtain the conditional distribution $\eta_i \mid e_i \sim N_{J-1}(\gamma \tilde e_i,\psi)$, where $\tilde e_i= \frac{e_i}{1-\delta_1^2}$. With that, we construct a multivariate linear regression model 
\begin{equation}
\eta_i = \gamma\tilde e_i + \zeta_i, \quad\quad \zeta_i\sim N_{J-1}(0,\psi).
\end{equation} 

Here, $\gamma$ and $\psi$ in Step $4$ are the parameters of the above multivariate regression, thus can be sampled with Gibbs using conjugate priors. With a normal prior on $\gamma$, we obtain the conditional posterior for $\gamma$
\begin{equation}
\gamma \mid \delta, w,\psi, Z,\beta,Y,X \sim N_{J-1}(\hat\gamma,\Sigma_\gamma),\quad \Sigma_\gamma=(\psi^{-1}\tilde e^\top \tilde e+ A_\gamma)^{-1}, \quad\hat\gamma=\Sigma_\gamma(\psi^{-1}\eta^\top \tilde e+A_\gamma B_\gamma),
\end{equation}
where $\tilde e = (\tilde e_1, \dots, \tilde e_n)^\top,$ and $\eta = (\eta_1^\top,\dots,\eta_n^\top)^\top$.

\begin{description}
    \item[\textbf{Step 5:}] Generate $\psi^{(r)} \sim p(\psi \mid \delta^{(r)}, w^{(r)},\gamma^{(r)}, Z^{(r-1)},\beta^{(r-1)},Y,X)$.
\end{description}    

The full conditional of $\psi$ can be obtained using the standard conjugate prior theory in the Bayesian normal linear model framework. Specifically, the conditional posterior distribution is also inverse-Wishart: 
\begin{equation}
\psi \mid \delta, w,\gamma, Z,\beta,Y,X \sim IW_{J-1}(J+3+N, V+\sum^N_{i=1}\zeta_{i}\zeta_{i}^\top ).
\end{equation}

\begin{description}
    \item[\textbf{Step 6:}] Generate $Z^{(r)} \sim p(Z \mid \delta^{(r)}, w^{(r)},\gamma^{(r)},\psi^{(r)},\beta^{(r-1)},Y,X)$.
\end{description}
As observations are assumed to be independent, we draw each of $Z_i$ separately. Then, we approximate the full conditional joint posterior of $Z_i$ by drawing sequentially from the $J$ conditional distributions of each element in $Z_i$ denoted as $z_{ij}$, conditioning on all other elements, as in \cite{mcculloch1994exact}. 

Moreover, note that each of the $z_{ij}$ comes from a univariate truncated normal distribution because $Z_i$ is constrained through $Y_i$ given the linear constraints in \eqref{eq:Y}. Each $z_{ij}$ is generated from
\begin{equation}
\begin{split}
z_{ij} \mid Z_{i(-j)},\beta,\delta, w_i,G ,Y_i,X_i\sim {N}^+_{max(Z_{i(-j)},0)}(m_{ij},\tau^2_{ij}), \quad \text{if} \quad Y_i=j, \\
z_{ij} \mid Z_{i(-j)},\beta,\delta, w_i,G ,Y_i,X_i\sim {N}^-_{max(Z_{i(-j)},0)}(m_{ij},\tau^2_{ij}), \quad \text{if} \quad Y_i\neq j,
\end{split}
\end{equation}
where $Z_{i(-j)}=(z_{i1},\dots,z_{ij-1},z_{ij+1},\dots,z_{iJ})$. By partitioning $G$ as
\begin{equation}
G=
\begin{bmatrix}
\Sigma_u{_{(-j)(-j)}} & \Sigma_u{_{(-j)j}}\\
\Sigma_u{_{j(-j)}}&  \Sigma_u{_{jj}} 
\end{bmatrix}^{-1},
\end{equation}
the conditional mean and variance are $m_{ij}=x^\top_{ij}\beta+\delta w_i+F^\top(Z_{i(-j)}-X_{i(-j)}\beta-\delta w_{i})$ and $\tau^2_{ij}=\Sigma_u{_{jj}}-\Sigma_u{_{j(-j)}}F$, respectively, where $F=\Sigma^{-1}_u{_{(-j)(-j)}}\Sigma_u{_{(-j)j}}$. Also, $x_{ij}$ refers to the $j$th row in $X_i$, while $X_{i(-j)}$ denotes $X_i$ with the $j$th row removed.

\begin{description}
    \item[\textbf{Step 7:}] Generate $\beta^{(r)} \sim p(\beta \mid \delta^{(r)}, w^{(r)},\gamma^{(r)},\psi^{(r)},Z^{(r)},Y,X)$.
\end{description}
By conjugacy, the conditional posterior for $\beta$ is also normal, that is, 
\begin{equation}
\beta \mid \delta, w,G, Z,X \sim N_k(\hat\beta,\Sigma_\beta),\quad \Sigma_\beta=(X^{*\top}X^*+A_\beta)^{-1}, \quad\hat\beta=\Sigma_\beta(X^{*\top}Z^*_\beta+A_\beta B_\beta),
\end{equation}
where $Z^*_\beta=\left(\left(Z_1^*-w_1^*\delta)^\top,\dots,(Z_N^*-w_N^*\delta\right)^\top\right)^\top$.

\clearpage
\section{Additional results numerical experiments}\label{A:numerical}

\begin{figure}[ht!]
\centering
\includegraphics[width=\textwidth]{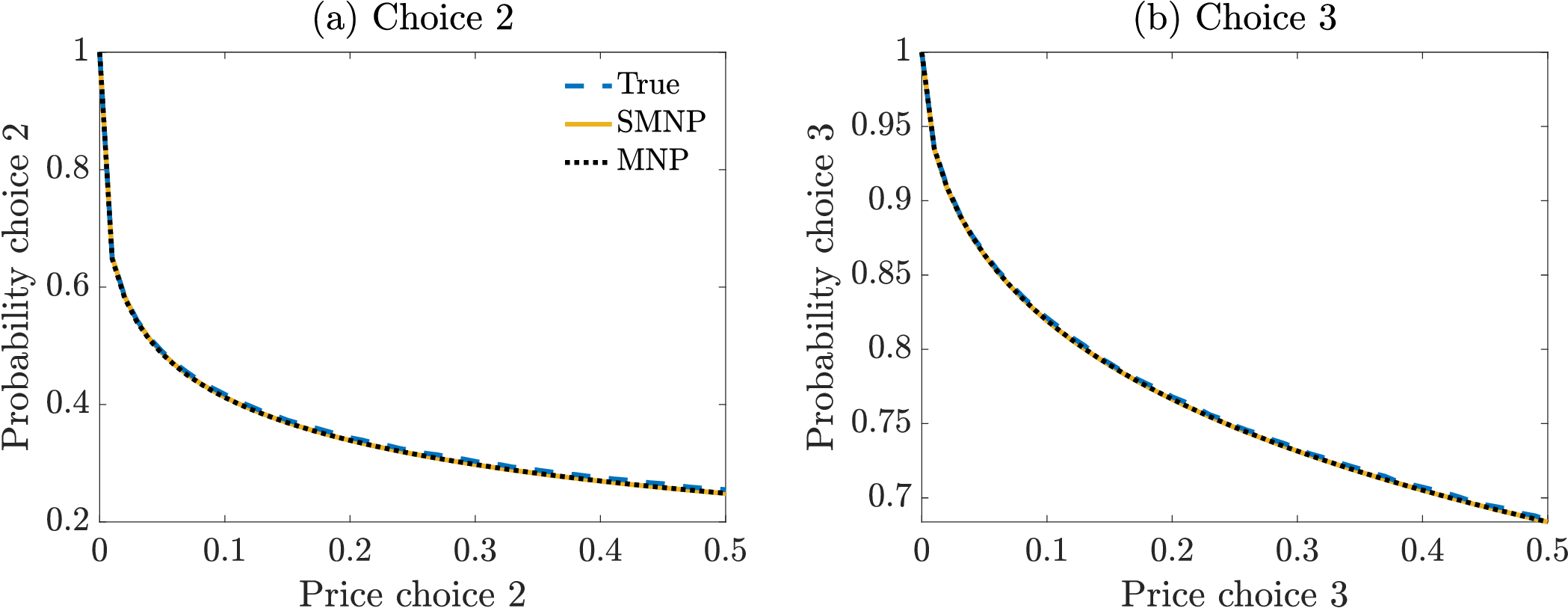}
\caption{Choice probabilities given different product prices indicated on the x-axis.  The solid yellow and dotted black lines represent the estimated choice probabilities using the SMNP and MNP models, respectively. The dashed blue line indicates the true choice probability under a MNP data generating process.}
\label{fig:choiceprob_normal}
\end{figure}

\clearpage
 
\section{Additional results empirical applications}\label{A:application}

\begin{figure}[ht!]
\centering
\includegraphics[width=\textwidth]{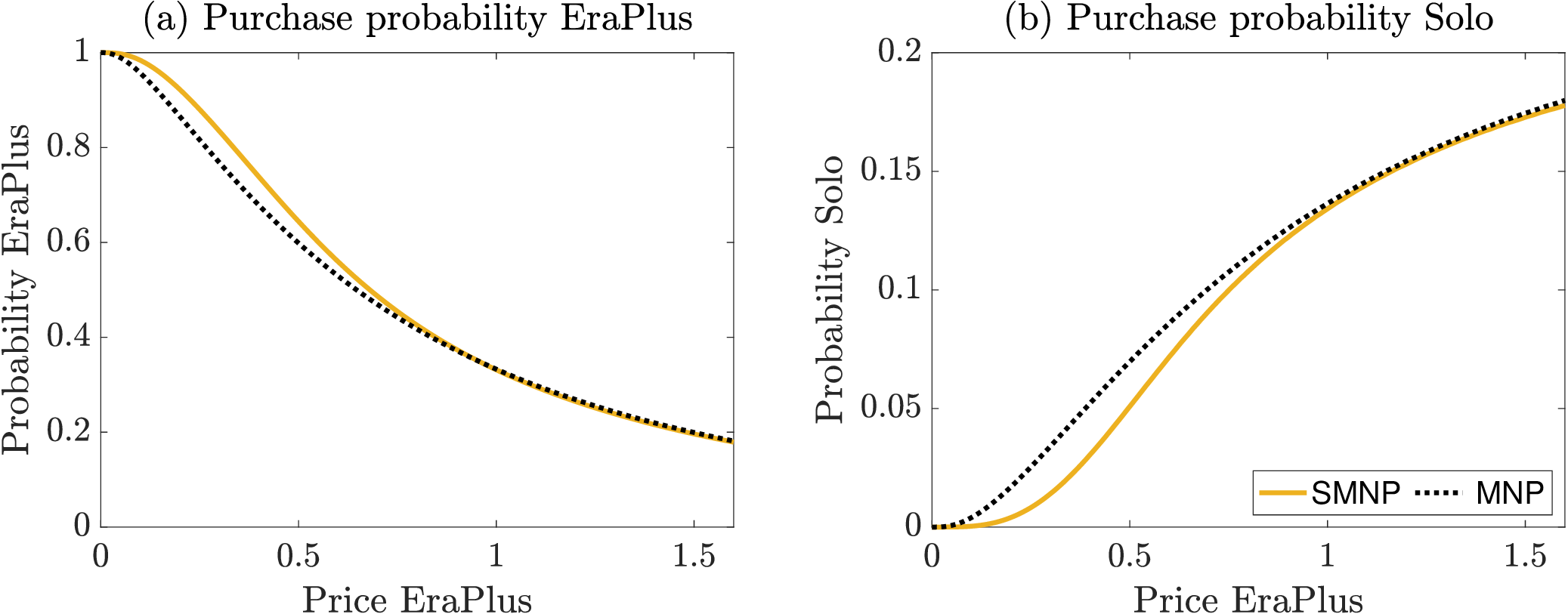}
\caption{Choice probabilities given the price of EraPlus indicated on the x-axis.  The solid yellow and dotted black lines represent the estimated choice probabilities using the SMNP and MNP models, respectively.}
\label{fig:choiceprob_detergent}
\end{figure}

\begin{figure}[ht!]
\centering
\includegraphics[width=\textwidth]{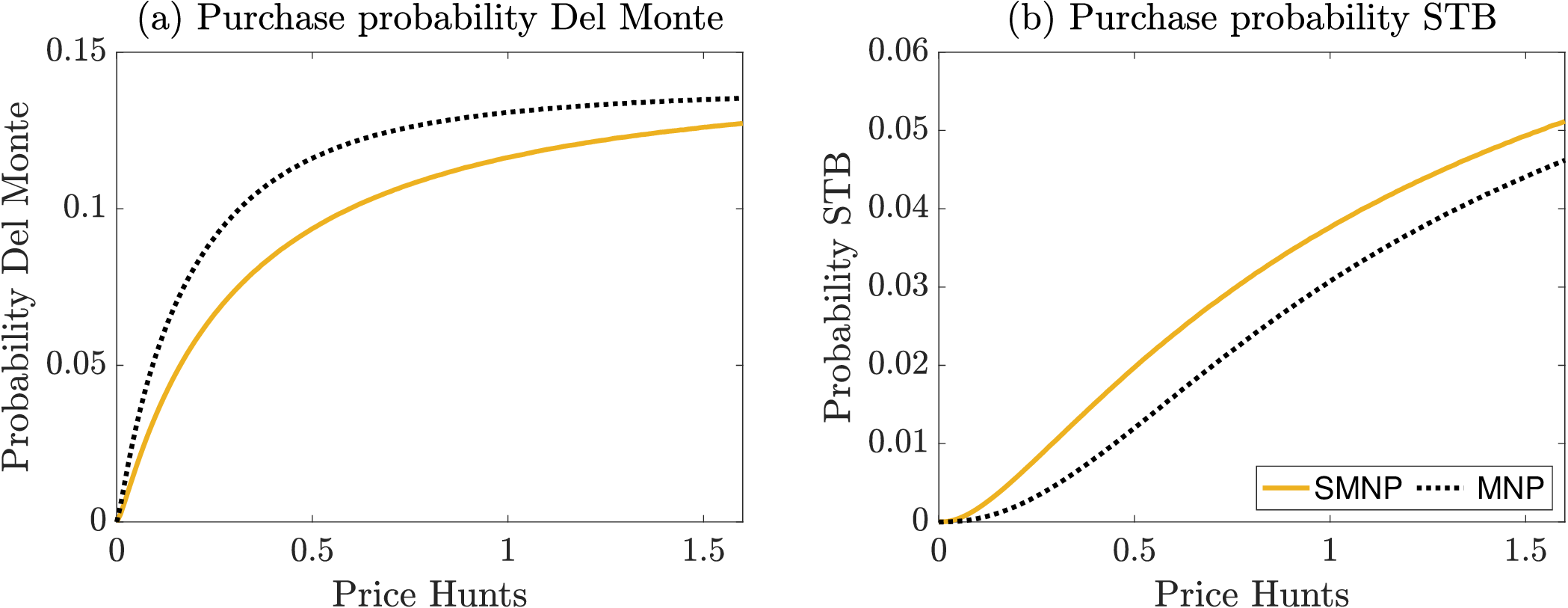}
\caption{Choice probabilities given the price of Hunts indicated on the x-axis.  The solid yellow and dotted black lines represent the estimated choice probabilities using the SMNP and MNP models, respectively.}
\label{fig:choiceprob_ketchup}
\end{figure}

\end{document}